\documentclass[lettersize,journal]{IEEEtran}
\usepackage{amsmath,amsfonts}
\usepackage{algorithmic}
\usepackage{algorithm}
\usepackage{array}
\usepackage[caption=false,font=normalsize,labelfont=sf,textfont=sf]{subfig}
\usepackage{textcomp}
\usepackage{stfloats}
\usepackage{url}
\usepackage{verbatim}
\usepackage{graphicx}
\usepackage{cite}
\usepackage{booktabs}
\usepackage{multirow}

\usepackage[numbers]{natbib}
\begin{document}

\title{Disentangling Prosody Representations \\with Unsupervised Speech Reconstruction}

\author{

Leyuan Qu, Taihao Li, \IEEEmembership{Member, IEEE}, Cornelius Weber, Theresa Pekarek-Rosin, \\ Fuji Ren, \IEEEmembership{Senior Member, IEEE} and Stefan Wermter, \IEEEmembership{Member, IEEE}


\thanks{This work was supported in part by the CML, LeCareBot and MoReSpace Projects funded by the DFG, the Major Scientific Project of Zhejiang Lab under Grant (No. 2020KB0AC01), and Youth Foundation Project of Zhejiang Lab (No. 111011-AA2301) \textit{(Corresponding author: Taihao Li).}}

\thanks{Leyuan Qu and Taihao Li are with the Institute of Artificial Intelligence, Zhejiang Lab, Hangzhou, China (e-mail: \{leyuan.qu, lith\}@zhejianglab.com).}

\thanks{Cornelius Weber, Theresa Pekarek-Rosin and Stefan Wermter are with the Department of Informatics, University of Hamburg, Hamburg, Germany (e-mail: \{cornelius.weber, theresa.pekarek-rosin, stefan.wermter\}@uni-hamburg.de).}

\thanks{Fuji Ren is with the School of Computer Science and Engineering, University of Electronic Science and Technology of China, Chengdu, China (e-mail: renfuji@uestc.edu.cn).}

}



\maketitle

\begin{abstract}
Human speech can be characterized by different components, including semantic content, speaker identity and prosodic information. Significant progress has been made in disentangling representations for semantic content and speaker identity in Automatic Speech Recognition (ASR) and speaker verification tasks respectively. However, it is still an open challenging research question to extract prosodic information because of the intrinsic association of different attributes, such as timbre and rhythm, and because of the need 
for supervised training schemes to achieve robust large-scale and speaker-independent ASR. The aim of this paper is to address the disentanglement of emotional prosody from speech based on unsupervised reconstruction. Specifically, we identify, design, implement and integrate three crucial components in our proposed speech reconstruction model Prosody2Vec: (1) a unit encoder that transforms speech signals into discrete units for semantic content, (2) a pretrained speaker verification model to generate speaker identity embeddings, and (3) a trainable prosody encoder to learn prosody representations. We first pretrain the Prosody2Vec representations on unlabelled emotional speech corpora, then fine-tune the model on specific datasets to perform Speech Emotion Recognition (SER) and Emotional Voice Conversion (EVC) tasks. Both objective (weighted and unweighted accuracies) and subjective (mean opinion score) evaluations on the EVC task suggest that Prosody2Vec effectively captures general prosodic features that can be smoothly transferred to other emotional speech. In addition, our SER experiments on the IEMOCAP dataset reveal that the prosody features learned by Prosody2Vec are complementary and beneficial for the performance of widely used speech pretraining models and surpass the state-of-the-art methods when combining Prosody2Vec with HuBERT representations. Some audio samples can be found on our demo website\footnote{https://leyuanqu.github.io/Prosody2Vec/}.
\end{abstract}

\begin{IEEEkeywords}
Prosody disentanglement, speech emotion recognition, emotional voice conversion.
\end{IEEEkeywords}

\section{Introduction}
\IEEEPARstart{H}{uman} speech contains rich information, which includes semantic content (what is spoken), speaker identity (who is speaking), and prosodic information (how is it spoken). 
Among them, prosody plays an important role in characterizing speaking styles, emotional states, and social intentions. Most importantly, humans express and perceive emotions
via various prosodic cues, for instance, sad speech often comes along with a low speaking rate, and angry emotion is usually accompanied by a raised pitch. According to Charles Darwin~\cite{darwin1998expression}, emotions are instinctive and present not only in humans in similar forms but also in many other species. Human infants can understand adults' emotions even without language skills~\cite{ruba2020development}. Therefore, enabling machines to capably recognize, understand and convey emotions is one of the crucial steps to achieving true artificial intelligence.


Learning meaningful prosodic representations has gained attention in recent years. Attention-enhanced Connectionist Temporal Classification (CTC)~\cite{zhao2019attention} and attention pooling~\cite{li2018attention} are utilized to dynamically capture useful temporal information for Speech Emotion Recognition (SER). Additionally, deep belief networks~\cite{luo2016emotional} and continuous wavelet transform~\cite{luo2017emotional} are utilized to learn prosodic features for Emotional Voice Conversion (EVC). However, model performance is greatly limited due to the lack of large-scale and high-quality emotional speech corpora.

Hence, disentangling prosodic information with unsupervised learning has been a promising direction, which includes Text-to-Speech (TTS) based style learning, such as automatically discovering expressive styles with global style tokens~\cite{wang2018style}. Moreover, an information bottleneck is used to control the information flow by careful design, such as in SpeechFlow~\cite{qian2020unsupervised}. In addition, mutual information loss is adopted to purify prosody representations, such as in a mutual information neural estimator~\cite{zhang2021estimating}. However, unsupervised methods usually require a well-trained Automatic Speech Recognition (ASR) system to decompose semantic content from speech. It is challenging to train a qualified ASR model with good performance, especially on emotional speech, since creating massive labeled corpora is time- and cost-consuming.

Another method is based on self-supervised learning by leveraging a large amount of unlabeled speech data. ~\citet{chen2022wavlm} propose WavLM and achieve state-of-the-art performance by fine-tuning the pretrained model on SER tasks. Nevertheless, the self-supervised learning models are mostly trained with mask prediction, similar to BERT~\cite{DevlinCLT19}, which leads the model to focus more on semantic content and local variations but neglect non-verbal and global information. Psychologists~\cite{schirmer2017emotion} found that the superior temporal gyrus\textemdash the site of the auditory association cortex\textemdash is more activated by longer audio, which reveals that humans tend to perceive emotions with long-term cues. Hence, it is critical to capture global or long-term prosodic changes.

The recent self-supervised model HuBERT~\cite{hsu2021hubert} integrates quantization into pretraining, where, instead of directly predicting the masked low-level acoustic features, HuBERT treats clustered Mel-Frequency Cepstral Coefficient (MFCC) features or clustered intermediate layer outputs by k-means as training targets. It has been proven that the quantization procedure can successfully filter non-verbal information, such as prosodic information~\cite{kreuk2021textless} and speaker identity~\cite{kharitonov2022textless}. Inspired by the above findings, we propose Prosody2Vec which does not require any human annotations and reconstructs emotional speech in an unsupervised fashion by conditioning on three information flows: (1) a unit encoder which is based on the pretrained HuBERT model to filter paralinguistic information and preserve only semantic content, (2) a pretrained Speaker Recognition (SR) model to generate speaker identity embeddings, and (3) a trainable prosody encoder to learn prosody representations.

Previous works, for instance, NANSY~\cite{choi2021neural} and SpeechFlow~\cite{qian2020unsupervised}, perform controllable or fine-grained speech synthesis by factorizing detailed prosodic attributes, such as pitch and rhythm. Instead of disentangling individual attributes, we aim to learn prosodic representations that reflect the combined effect of different prosodic attributes. Additionally, current speech representation models, e.g. HuBERT, focus more on local semantics modeling on a millisecond time scale, which results in an incapacity to represent long-term information. However, the production and perception of emotion usually require a relatively long, second-level time scale~\cite{bachorowski1999vocal}.

In addition, previous supervised work using Variational Autoencoder (VAE)~\cite{zhang2019learning} and Vector-Quantize VAE (VQ-VAE)~\cite{wang2022unsupervised} requires human annotations (text transcriptions) to provide semantic content. 
The lack of large-scale labeled emotional or expressive datasets significantly restricts model performance.
In comparison, the proposed Prosody2Vec model leverages self-supervised pretraining, quantization, and refinement schemes to represent semantics without
text annotations, which enables Prosody2Vec to train with large-scale datasets containing variant speaker styles.

Comparing with unsupervised methods, like AutoVC~\cite{qian2019autovc} and SpeechFlow~\cite{qian2020unsupervised}, which control information flows by several carefully designed bottleneck autoencoder modules.
It is complicated and time-consuming to balance different information flows and determine a suitable dimension through trial and error. However, we explicitly provide semantic and speaker information by pretrained models in Prosody2Vec. Only the prosody encoder needs to be controlled and tuned.

In this paper, our goal is to capture global or utterance-level variations, which are complementary to the semantic representations learned by speech representation models. The main contributions of this paper are:

\begin{enumerate}
\item We propose a novel model, Prosody2Vec, to learn prosody information from speech, which requires neither emotion labels nor transcribed speech for robust ASR system building. 
\item The SER results on the IEMOCAP dataset reveal that, after pretraining with large-scale unlabelled data, Prosody2Vec can successfully capture prosody variations, which is complementary to the widely used speech pretraining models, such as Wav2Vec2~\cite{baevski2020wav2vec} and HuBERT. We surpass the state-of-the-art method when combining Prosody2Vec with HuBERT.

\item We conduct subjective and objective evaluations on EVC tasks. The experimental results demonstrate that Prosody2Vec can effectively convert a given emotional reference into any speech utterance.
\end{enumerate}

The rest of the paper is organized as follows. Section~\ref{related-work} reviews some related work on  prosody disentanglement, SER and EVC. Section~\ref{architecture} details the proposed Prosody2Vec architecture. We introduce the used datasets, Prosody2Vec pretraining, and evaluate our proposed method on SER, and EVC tasks in Section~\ref{experiment}. 
We conduct a series of ablation studies to deelp understand the Prosody2Vec model in Section~\ref{sec:ablation}.
Some potential applications, such as, zero-shot emotional, speaking, and singing style transfer, are presented in Section~\ref{discussion}. We conclude and summarize the results of this paper in Section~\ref{conclusion}.

\section{Existing Research Methods}
\label{related-work}
In this section, we briefly review related work on prosody disentanglement, speech emotion recognition, and emotional voice conversion.

\subsection{Prosody Disentanglement}

Prosody disentanglement aims to decompose different acoustic or phonetic speech attributes, such as pitch, timbre, rhythm, intonation, loudness, and tempo. Current approaches can be mainly divided into three parts: (1) TTS-based style learning, (2) information bottleneck~\cite{tishby2000information}, and (3) mutual information loss. 

TTS-based methods force additional attribute encoders to provide prosodic information when transforming text sequences into speech signals. ~\citet{skerry2018towards} integrate an encoder module into the Tacotron~\cite{skerry2018towards} TTS system to capture meaningful variations of prosody and successfully perform speaking style transfer. Subsequently, ~\citet{wang2018style} introduce ``global style tokens" to automatically discover expressive styles. In addition, Variational Autoencoder (VAE)~\cite{zhang2019learning} and Vector-Quantize VAE (VQ-VAE)~\cite{wang2022unsupervised} are adopted to learn continual and discretized prosody representations from a reference audio respectively.

The basic idea of information bottleneck approaches is to control the information flow by carefully designing appropriate bottlenecks. AutoVC~\cite{qian2019autovc} adopts a properly tuned autoencoder as the information bottleneck to force the model to disentangle linguistic content and speaker identity with self-reconstruction. SpeechFlow~\cite{qian2020unsupervised} extends the AutoVC model by constraining the dimension of representations and adding randomly sampled noise to blindly split content, pitch, timbre, and rhythm from speech. However, bottlenecks need to be carefully designed and are sensitive to the dimension of latent space.


The use of mutual information loss is minimizing information redundancy between different attributes. To allow more precise control over different speech attributes, ~\citet{kumar2021learning} formulate a modified Variational Auto-Encoder (VAE) loss function to penalize the similarity between different attribute representations. ~\citet{weston2021learning} introduce a self-supervised contrastive model that adopts product quantization to disentangle non-timbral prosodic information from raw audio.

Different from the aforementioned approach, in this paper, we aim to disentangle a global prosody representation from speech instead of factorizing detailed attributes for controllable or fine-grained speech synthesis. Different from speech synthesis, emotion recognition, and conversion tasks rely more on prosody representations that reflect the combined effect of different prosodic attributes.

\subsection{Speech Emotion Recognition}

In this paper, we only review the recent work on categorical emotion classification. Advanced models and methods are proposed to overcome the bottleneck caused by limited emotional speech corpora. ~\citet{LakomkinWMW17} utilize fine-tuning methods and progressive networks~\cite{rusu2016progressive} to transfer ASR representations to emotion classification. In addition, attention-enhanced Connectionist Temporal Classification (CTC)~\cite{zhao2019attention} and attention pooling~\cite{li2018attention} are utilized to dynamically weigh the contribution of temporal changes in an utterance. Furthermore, different multi-task architectures are designed to learn more generalized features. For instance, building SER models with both discrete and continual labels~\cite{le2017discretized}, integrating naturalness prediction as an auxiliary task~\cite{atmaja2022speech}, and exploiting secondary emotion labels by the perceptual evaluation of annotators after aggregating them~\cite{chou2022exploiting}. 

Inspired by the success of self-supervised pretraining in ASR tasks, researchers directly utilize pretrained speech representations for SER, such as attempting different fine-tuning strategies~\cite{wang2021fine}. However, modern speech representation models focus more on local variations or semantic information but rarely take emotional or prosodic cues into account. In this paper, we propose to adopt unsupervised pretraining to capture global prosodic information at an utterance level, which is complementary to the widely used speech representation models, such as Wav2Vec2 and HuBERT.

\begin{figure*}[!th]
\setlength{\abovecaptionskip}{0cm}
\setlength{\belowcaptionskip}{-0cm}
\centering 
\small
\includegraphics[width=0.7\textwidth]{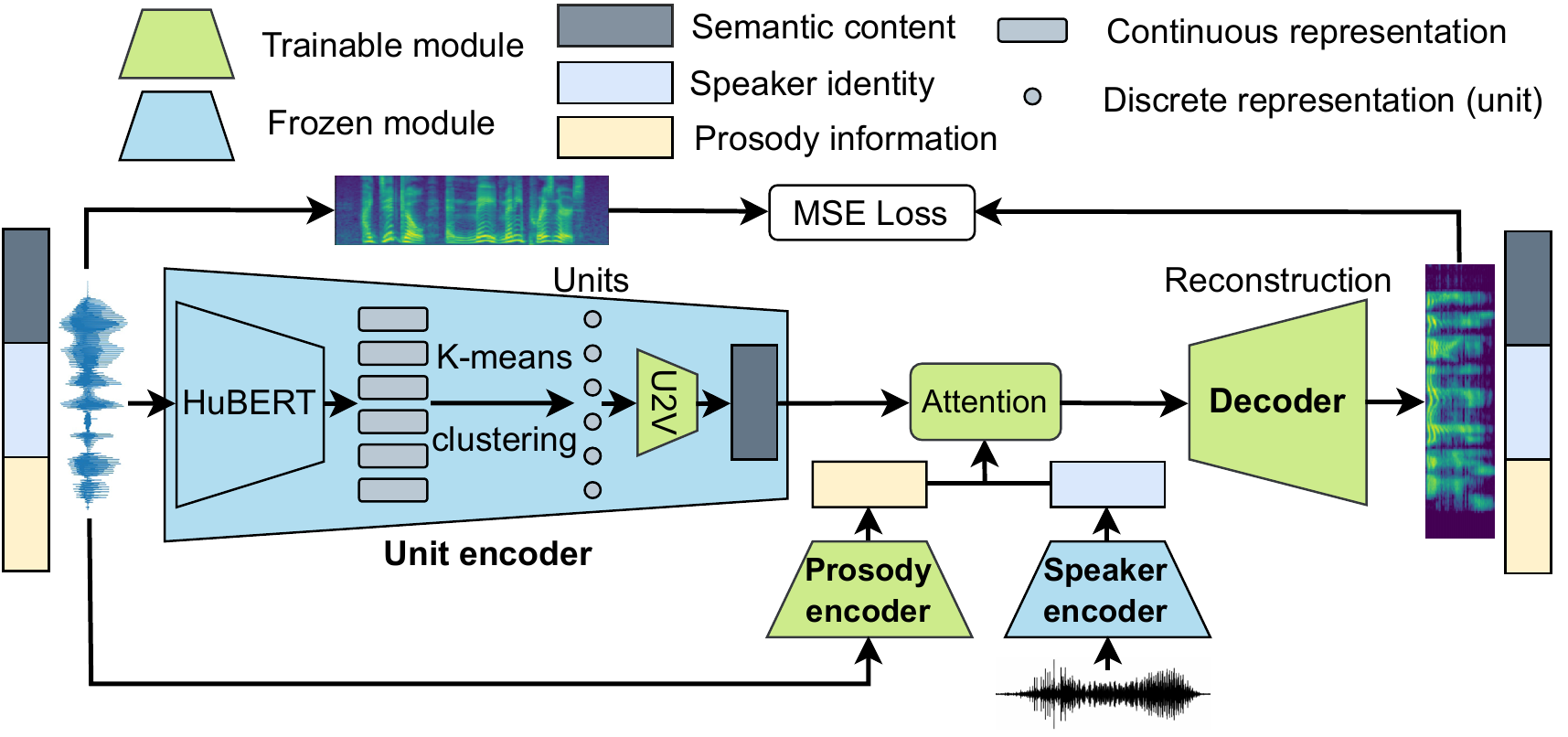}
\caption{The architecture of Prosody2Vec. During training, the weights in U2V, prosody encoder, attention module and the decoder are updated, while the HuBERT representations and the k-means algorithm in the unit encoder are performed beforehand, and the speaker encoder is frozen. The model receives two different mel-spectrograms as inputs and aims to reconstruct a mel-spectrogram similar to the one fed into the unit encoder.}

\label{model}
\vspace{-0.5cm}
\end{figure*}

\subsection{Emotional Voice Conversion}

The EVC task aims to convert a speaker’s speech from one emotion to another while preserving semantic contents and speaker identities. Typically, parallel data is required to perform frame-to-frame mapping. ~\citet{csicsman2017transformation} utilize continuous wavelet transforms to map source and target audios on the side of F0, energy contour, and duration. Subsequently, deep belief networks and deep neural networks are used to build mel-cepstral coefficients and F0 mappings respectively~\cite{luo2016emotional}. Frame-to-frame methods assume the same utterance length between input and generated speech. However, different emotions are conveyed with various segments or syllable duration, and it is unreasonable to restrict different emotional speech utterances to have the same duration. In addition, collecting parallel emotional datasets is expensive and time demanding.

To tackle the above issues, different models using nonparallel data are thereby proposed. For instance, Cycle Generative Adversarial Networks (GANs)~\cite{zhou2020transforming} and StarGANs~\cite{rizos2020stargan} are used to predict spectrum and prosody mappings. Besides, ~\citet{zhou2021limited} propose a sequence-to-sequence framework, in which TTS and SER tasks are jointly trained with EVC.  ~\citet{zhou2022emotion} propose Emovox to control fine-grained emotional intensity by integrating intensity and emotion classification into EVC training. Inspired by the mechanism of speech production, ~\citet{luo2022decoupling} design a source-filter network to learn speaker-independent emotional features.
Nonetheless, these systems usually rely on additional annotations, such as emotion labels, text transcriptions, and speech intensities. Different from the current EVC methods, we conduct EVC experiments with unsupervised emotional speech reconstruction, which requires neither paired speech nor additional labels.

\section{Prosody2Vec Architecture}
\label{architecture}
To leverage disentangled semantic content by the quantization procedure in HuBERT, we propose Prosody2Vec, as shown in Fig.~\ref{model}, which consists of four crucial modules: a unit encoder, a speaker encoder, a prosody encoder, and a decoder. We detail each module in the following subsections.

\subsection{Unit Encoder}

As shown in Fig.~\ref{model}, the unit encoder firstly extracts latent representations from original speech signals with pretrained HuBERT. Then, the k-means algorithm and deduplication process are used for vector quantization and semantics refinement respectively. The process of quantization and refinement can effectively remove the speaker and prosody information from the original speech, which is discussed in Section~\ref{sec:ablation}.
Lastly, a Unit2Vec (U2V) module transforms the deduplicated discrete units into latent space for model training. We detail the quantization and refinement procedures as follows.

\subsubsection{Quantization}

The backbone of the unit encoder is based on the recent self-supervised model HuBERT which learns speech representations by predicting masked parts, similar to BERT~\cite{DevlinCLT19}. HuBERT\footnote{https://huggingface.co/facebook/hubert-base-ls960} is pretrained on the LibriSpeech~\cite{panayotov2015librispeech} dataset with 960 hours data. We first extract dense representations at the frame level for each utterance from waveform signals.

We denote a sequence of waveform signals as $x=(x_1,...,x_T)$, where $T$ is the length of an audio waveform. The audio sample $x$ is transformed into a sequence of continuous vectors by the pretrained HuBERT:

\begin{equation}
\setlength{\abovedisplayskip}{1pt}
\setlength{\belowdisplayskip}{1pt}
y = HuBERT(x)
\label{attention1}
\end{equation}

\noindent with $y=(y_1,...,y_L)$, where $L < T$. The dense representation $y$ is often used for downstream tasks, e.g. ASR and SER.

Different from previous work, we quantize continuous vectors into discrete units to filter speaker information and refine semantic content.
The quantization procedure can be performed by the k-means algorithm on the dense representations:

\begin{equation}
\setlength{\abovedisplayskip}{1pt}
\setlength{\belowdisplayskip}{1pt}
u = k\mbox{-}means(y)
\label{attention2}
\end{equation}

\noindent with $u=(u_1,...,u_L)$ and $u_i \in \left\{1,N\right\}$, where $N$ is the number of clusters. 
The dense representations embedded by HuBERT are quantized into discrete units (cluster labels) $u$ frame by frame, e.g. "23, 23, 23, 2, 2, ..., 57".

\subsubsection{Refinement}

Subsequently, to refine the quantized sequences, we perform a refinement procedure since the adjacent repetitions may carry duration and rhythm information.
Specifically, we deduplicate the unit sequence $u$ to $\widetilde{u}$ by merging and removing repetitions, e.g. "23, 23, 23, 2, 2, ..., 57" $\rightarrow$ "23, 2, ..., 57", which purifies the speech units and avoids the leak of prosody information. As a consequence, Prosody2Vec can only capture rhythm and duration information from the prosody encoder.
We hereafter use \textbf{speech units} to represent the deduplicated discrete units and refer to $N$ as vocabulary size.
The purified speech units are utilized to represent semantic content.
Table~\ref{table:discrete-units} shows the discrete speech units of a random utterance with a vocabulary size of 50, 100, and 200 units.

\subsubsection{Unit2Vec}
The Unit2Vec (U2V) module maps the discrete speech units to a continuous latent space with an embedding layer, followed by three 1D-CNN layers and one bi-directional Long Short-Term Memory (LSTM) layer. The detailed configurations are listed in Table~\ref{tab:model_cofiguration}.

\begin{table}[th]
\centering
\caption{Speech units for the utterance of ``I'M DAMN GOOD AT MY JOB", where VS is short for vocabulary size.}
\begin{tabular}{cp{7cm}} 
\toprule
VS &
  \multicolumn{1}{c}{Units} \\ \hline
50 &
  0 2 15 20 18 0 8 3 27 28 7 46 7 37 20 49 47 45 4 43 31 3 28 27 28 46 37 49 45 41 19 0 31 26 47 35 44 27 40 43 4 7 44 49 25 47 45 0 8 3 46 20 \\ \hline
100 &
  71 39 67 54 57 86 68 16 18 66 27 57 31 45 64 53 38 16 50 18 66 78 90 69 90 35 53 9 85 53 73 74 2 50 24 58 32 64 1 66 27 21 98 87 24 17 24 61 24 61 43 16 20 \\ \hline
200 &
  14 131 161 42 11 117 110 145 5 155 53 93 156 13 30 156 89 86 144 50 28 113 25 53 93 66 156 146 178 91 58 187 69 127 163 70 177 106 145 108 184 13 156 195 171 98 28 16 26 97 83 155 79 92 \\
\bottomrule
\end{tabular}
\label{table:discrete-units}
\vspace{-0.3cm}
\end{table}
 
\subsection{Speaker Encoder and Prosody Encoder}

\begin{figure}[t]
  \setlength{\abovecaptionskip}{0cm}
  \setlength{\belowcaptionskip}{-0cm}
  \centering
  \includegraphics[width=7cm]{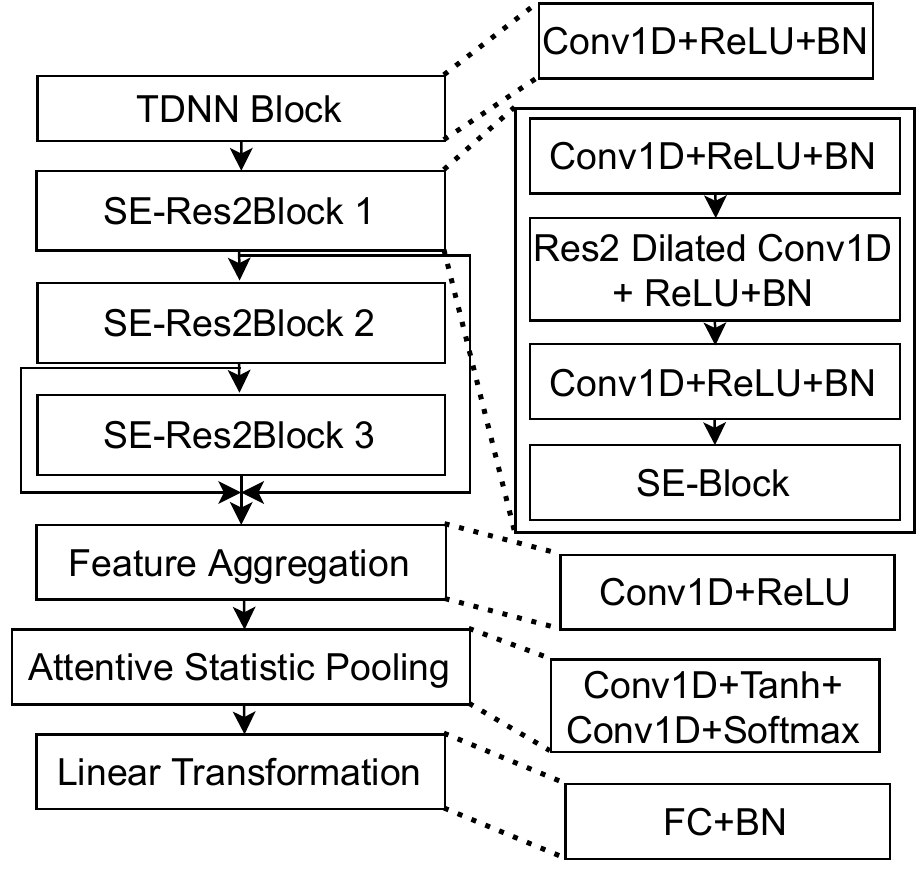}  
  \caption{Architecture of ECAPA-TDNN, where SE is short for squeeze excitation.}
  \label{fig:ecapa}
\vspace{-0.3cm}
\end{figure}

The speaker encoder is based on the ECAPA-TDNN~\cite{DesplanquesTD20} speaker verification model~\cite{DesplanquesTD20}, which is pretrained on the VoxCeleb2~\cite{chung2018voxceleb2} dataset and achieves state-of-the-art results with a 0.87\% equal error rate. We show the ECAPA-TDNN~\cite{DesplanquesTD20} details in Fig.~\ref{fig:ecapa}, which begins with a Time Delay Neural Network (TDNN)~\cite{PeddintiPK15} layer, followed by three SE-Res2Blocks. Each SE-Res2Block consists of 2 1D-CNN layers, a dilated Res2Net~\cite{gao2019res2net} and a Squeeze-Excitation (SE)~\cite{hu2018squeeze} block. Then a 1D-CNN combines outputs from the three previous SE-Res2Blocks, followed by attentive statistics pooling and a Fully Connected (FC) layer. The dilation factors used in the first three SE-Res2Blocks are 2, 3, and 4 respectively. The channel size used in the above three blocks is 1024 with a kernel size of 3.

The output vectors with 192 dimensions from the last FC layer of a model pretrained on the Voxceleb2 dataset are used as the speaker embeddings. In case the decoder directly learns prosodic information from speaker embeddings, we input a different audio belonging to the same speaker to the speaker encoder during training. 
 The HuBERT representations and k-means algorithm in the unit encoder are performed beforehand. 
 During training, the weights in U2V, prosody encoder, attention module and the decoder are updated, while the speaker encoder are frozen.
 The dense representations of speech signals are extracted by pretrained HuBERT beforehand and speech units quantized with k-means are saved locally.
 We freeze the pretrained speaker encoder to maintain the knowledge learned on the big Voxceleb2 dataset and ensure only speaker-related information is delivered.

The architecture of the prosody encoder is also based on ECAPA-TDNN, which is the same as the speaker encoder, but with random initialization. The weights 
 of prosody encoder are updated by minimizing the mean square error (MSE) between the generated and original mel-spectrograms. The prosody encoder is fed with the same mel-spectrograms as the one used in the unit encoder.

\subsection{Decoder}
Our decoder is similar to the one used in Tacotron2~\cite{shen2018natural}. The decoder reconstructs mel-spectrograms utilizing the outputs from the aforementioned three encoders. A location-aware attention mechanism~\cite{chorowski2015attention} is used to bridge the encoders and the decoder. The decoder consists of one unidirectional LSTM layer followed by one linear projection layer to map the intermediate representations to the dimension of the mel-scale filter bank. In addition, two FC layers (PreNet) are used to embed the ground-truth mel-spectrograms into a latent space.

Table~\ref{tab:model_cofiguration} shows the configuration of U2V, attention module, and decoder. More details about the location-aware attention mechanism can be found in the approach by ~\citet{chorowski2015attention} and LipSound2~\cite{lipspund2qu}.

\begin{table}[th]
  \setlength{\abovecaptionskip}{0cm}
  \setlength{\belowcaptionskip}{-0cm}
  \caption{Configuration of U2V, attention and decoder of Prosody2Vec.}
  \label{tab:model_cofiguration}
  \centering
  \setlength\tabcolsep{2pt}%
  \resizebox{9cm}{!}{
    \begin{tabular}{ccccc}
    \toprule
    \textbf{Layer} & \textbf{Kernel} & \textbf{Stride} & \textbf{Padding} & \textbf{Channels/Nodes}\\
    \midrule
    \textbf{U2V} \\
    Conv1D 1                   & 5         & 1          & 2                     & 512                          \\
    Conv1D 2                   & 5            & 1           & 2                        & 512                           \\
    Conv1D 3                   & 5            & 1           & 2                        & 512                           \\
    BiLSTM                & -           & -           & -                         & 256                           \\
    \midrule
    \textbf{Attention} \\
    Attention LSTM & -           & -           & - & 1408\\
    Query FC & -           & -           & - & 128\\
    Memory FC& -           & -           & - & 128\\
    Location Conv1D             & 31           & 1          & 15  & 32\\
    Location FC& -           & -           & - & 128\\
    Weight FC & -           & -           & - & 1\\
    \midrule
    \textbf{Decoder}\\
    PreNet FC 1 & -           & -           & - & 256\\
    PreNet FC 2 & -           & -           & - & 256\\
    Decoder LSTM & -           & -           & - & 1024\\
    Linear Projection FC& -           & -           & - & 80\\
    \bottomrule
    \end{tabular}}
\vspace{-0.3cm}
\end{table}

\section{Experiments}
\label{experiment}
In this section, we describe the setup and datasets used for the pretraining Prosody2Vec. We conduct comprehensive assessments and report results for SER and EVC experiments.

\subsection{Datasets}
We use spontaneous and emotional speech datasets, i.e. LRS3-TED~\cite{afouras2018lrs3}, MSP-PODCAST~\cite{Lotfian_2019_3}, MSP-IMPROV~\cite{busso2016msp} and, OMG~\cite{barros2018omg} datasets, to pretrain the proposed model, then fine-tune it on IEMOCAP~\cite{busso2008iemocap} and ESD~\cite{zhou2022emotional} datasets to perform SER and EVC experiments respectively. The statistics of all datasets used in this paper are shown in Table~\ref{table:datasets}.

\begin{itemize}
\item \textbf{LRS3-TED}~\cite{afouras2018lrs3}: an audio-visual dataset collected from TED and TEDx talks with spontaneous speech and various speaking styles and emotions. It is comprised of over 400 hours of video by more than 5000 speakers and contains an extensive vocabulary.
\item \textbf{MSP-PODCAST}~\cite{Lotfian_2019_3}: a large real-scenario dataset including extensive emotional speech from podcast recordings. It contains speech about various topics, such as movies, politics, and sports.
\item \textbf{MSP-IMPROV}~\cite{busso2016msp}:  a multimodal dataset recorded in spontaneous dyadic interactions in which the emotions are evoked by an elicitation scheme.
\item \textbf{OMG}~\cite{barros2018omg}: an audio-visual dataset collected from YouTube with restricted keywords, for instance, “monologue”. The dataset allows the exploration of the long-term emotional behavior categorization by using contextual information.
\item \textbf{IEMOCAP}~\cite{busso2008iemocap}: a multimodal dataset recorded with elicited emotions by 10 actors in a fictitious scenario. The dataset provides audio and visual modalities, and motion information on the head, face, and hands during communication.
\item \textbf{ESD}~\cite{zhou2022emotional}:  an audio dataset with parallel emotional speech, in which actors are required to act 5 different emotions with the same text content. 
\end{itemize}


\begin{table}[th]
\centering
\caption{Overview of all corpora used in this paper. Spk: Speakers. Utt: Utterances.}
  \resizebox{8cm}{!}{
\begin{tabular}{ccccc}
\toprule
Dataset     & \#Spk. & \#Utt. & \#hours & Usage \\ \hline
LRS3-TED & 5090 & 151k & 437 & \multirow{4}{*}{\begin{tabular}[c]{@{}c@{}}Prosody2Vec\\ pretraining\end{tabular}} \\
MSP-PODCAST & 1285   & 62k    & 100     &       \\
MSP-IMPROV  & 12     & 8k     & 9.5     &       \\
OMG         & $\sim$500    & $\sim$7.4k   & 15      &       \\ \hline
IEMOCAP     & 10     & 10k    & 12.5    & SER   \\ \hline
ESD         & 20     & 35k    & 29      & EVC  \\
\bottomrule
\end{tabular}}
\label{table:datasets}
\vspace{-0.5cm}
\end{table}

\begin{table*}[!t]
\renewcommand\arraystretch{1.2}
\setlength{\abovecaptionskip}{0.cm}
\setlength{\belowcaptionskip}{-0.cm}
	\caption{\small Weighted Accuracy (WA$\uparrow$) of speech emotion recognition with leave-one-session-out settings. Note: results of Prosody2Vec, Wav2Vec2, and HuBERT listed in the table are fine-tuned on the IEMOCAP dataset, where the difference between single and fusion is whether it is combined with Prosody2Vec. Model fusion is performed by vector concatenation before use for classification by the last FC layer.}
	\label{tab:fusion-experiments}
	\vspace{0.8mm}
	\centering
	\small
	\setlength\tabcolsep{8pt}
		\resizebox{17cm}{!}{
\begin{tabular}{ccccccccccc}
\toprule
\multirow{2}{*}{Vocabulary Size} &
  \multirow{2}{*}{FBANK} &
  \multirow{2}{*}{Prosody2Vec} &
  \multicolumn{2}{c}{Wav2Vec2 Base} &
  \multicolumn{2}{c}{Wav2Vec2 Large} &
  \multicolumn{2}{c}{HuBERT Base} &
  \multicolumn{2}{c}{HuBERT Large} \\ \cline{4-11}
    &  &       & Single & Fusion & Single & Fusion & Single & Fusion & Single & Fusion \\ \hline
50 &
  \multirow{3}{*}{55.67} &
  63.24 &
  \multirow{3}{*}{66.62} &
  67.12 &
  \multirow{3}{*}{69.24} &
  70.27 &
  \multirow{3}{*}{66.88} &
  67.10 &
  \multirow{3}{*}{70.44} &
  70.88 \\ \cline{1-1} \cline{3-3} \cline{5-5} \cline{7-7} \cline{9-9} \cline{11-11} 
100 &  & 63.40 &        & 67.56  &        & 70.57  &        & 69.03  &        & 71.54  \\ \cline{1-1} \cline{3-3} \cline{5-5} \cline{7-7} \cline{9-9} \cline{11-11} 
200 &  & 64.10  &        & 69.14  &        & 71.21  &        & 69.40   &        & 72.42  \\
\bottomrule
\end{tabular}
		}
\vspace{-0.5cm}
\end{table*}

\subsection{Prosody2Vec Pretraining}

We merge the LRS3-TED, MSP-PODCAST, MSP-IMPROV, and OMG datasets for pretraining. 500 randomly selected samples from the above datasets are utilized for validation. We augment training data by perturbing speed with the factors of 0.9, 1.0, and 1.1. Furthermore, SpecAugment~\cite{park2019specaugment} with two frequency masks (maximum width of 50) is utilized on the fly during training. In addition, gradient clipping with a threshold of 1.0, early stopping, and scheduled sampling~\cite{bengio2015scheduled} are adopted to avoid overfitting. The Prosody2Vec model is pretrained with a batch size of 30 and 3000 warm-up steps. We use the Adam optimizer~\cite{kingma2014adam} and the cosine Learning Rate (LR) decay strategy with an initial value of 1e-3. The experiments are conducted on two 32G memory NVIDIA Tesla V100 GPUs in parallel. We pretrain three models with a vocabulary size of 50, 100, and 200 units to explore the effect of quantization. The entire pretraining procedure takes around three weeks for each model.

We extract the magnitude using the Short Time Fourier Transform (STFT) with 1024 frequency bins and a 64ms window size with a 16ms stride. The mel-scale spectrograms are obtained by applying an 80-channel mel filter bank to the magnitude. The model is optimized with Mean Squared Error (MSE) loss to minimize the distance between the generated and original mel-spectrograms.

\subsection{Experiments of Speech Emotion Recognition}
\subsubsection{Experimental Setups} The SER experiments are conducted on the widely used IEMOCAP dataset. We merge ``happy” and ``excited” into the category of ``happy” to balance each class. Finally, 5531 utterances are used for training and testing, which include four emotions, i.e. angry, sad, happy, and neutral. The dataset is comprised of five sessions with two speakers in each session. We conduct SER experiments with the following two settings to provide a comprehensive comparison with previous work~\cite{yang2021superb}:
\begin{itemize}
    \item \textbf{Leave-one-session-out} is performed with 5-fold cross-validation. In each round, one session is used for testing and another random session is used as a validation set. The remaining three sessions are treated as the training set.
    \item \textbf{Leave-one-speaker-out} means using one speaker for testing in one session and the other speaker in the same session is utilized for validation. Therefore, 10-fold cross-validation is performed.
\end{itemize}

We fine-tune the pretrained prosody encoder with one additional FC layer to perform emotion classification, in which LRs of 1e-4 and 5e-4 are used for the pretrained prosody encoder and for the last FC layer respectively.
The fusion experiments are conducted by concatenating the representations generated by the prosody encoder with the outputs of Wav2Vec2 or HuBERT. Then the concatenated vectors are fed into one FC layer for classification.

\subsubsection{Evaluation Metrics} We utilize the following two metrics to assess the Prosody2Vec performance on SER tasks.
\begin{itemize}
    \item \textbf{Weighted Accuracy (WA):} the accuracy of all utterances in the test set.
\end{itemize}

\begin{equation}
\setlength{\abovedisplayskip}{1pt}
\setlength{\belowdisplayskip}{1pt}
WA=\frac{\sum_{i=1}^M{U_i}}{N}
\label{eq:wa}
\end{equation}

\begin{itemize}
    \item \textbf{Unweighted Accuracy (UA):} the average accuracy of each emotion class.
\end{itemize}

\begin{equation}
\setlength{\abovedisplayskip}{1pt}
\setlength{\belowdisplayskip}{1pt}
UA=\frac{\sum_{i=1}^M{U_i/T_i}}{M}
\label{eq:ua}
\end{equation}

\noindent where $M$ and $N$ represent the number of emotion classes and the total number of utterances in the test set respectively. $U_i$ denotes the number of utterances with a correct prediction of the emotion class $i$ and $T_i$ is the total number of utterances of emotion class $i$.
    
\subsubsection{Experimental Results of Speech Emotion Recognition}

We compare the performance of using only acoustic FBANK features, only our pretrained Prosody2Vec, and only pretrained speech representation models, i.e. Wav2Vec2 and HuBERT, where the base and large models are trained on 960h LibriSpeech and 60kh Libri-light~\cite{kahn2020libri} respectively. In addition, we also report the results of combining Prosody2Vec with Wav2Vec2 or HuBERT. As shown in Table~\ref{tab:fusion-experiments}, Prosody2Vec surpasses the baseline model using FBANK features but is not as good as Wav2Vec2 or HuBERT. One reason is that Wav2Vec2 and HuBERT are trained with larger datasets, 960h or 60kh, whereas our model is trained on only 460h of speech data. Another potential reason is that the representations captured by the prosody encoder are more related to prosodic variations. In comparison to prosody information, semantic content learned by Wav2Vec2 or HuBERT is important for emotion recognition as well, which is also found in psychology~\cite{pell2011emotional}. Further improvement can be obtained when combining Prosody2Vec with Wav2Vec2 or HuBERT. Moreover, it seems that a bigger vocabulary size equals better performance.
Hence, we only report the results of vocabulary size 200 in the rest of the paper.

\begin{figure*}[h]
\setlength{\abovecaptionskip}{0cm}
\setlength{\belowcaptionskip}{-0cm}
\centering 
\small
\includegraphics[width=1\textwidth]{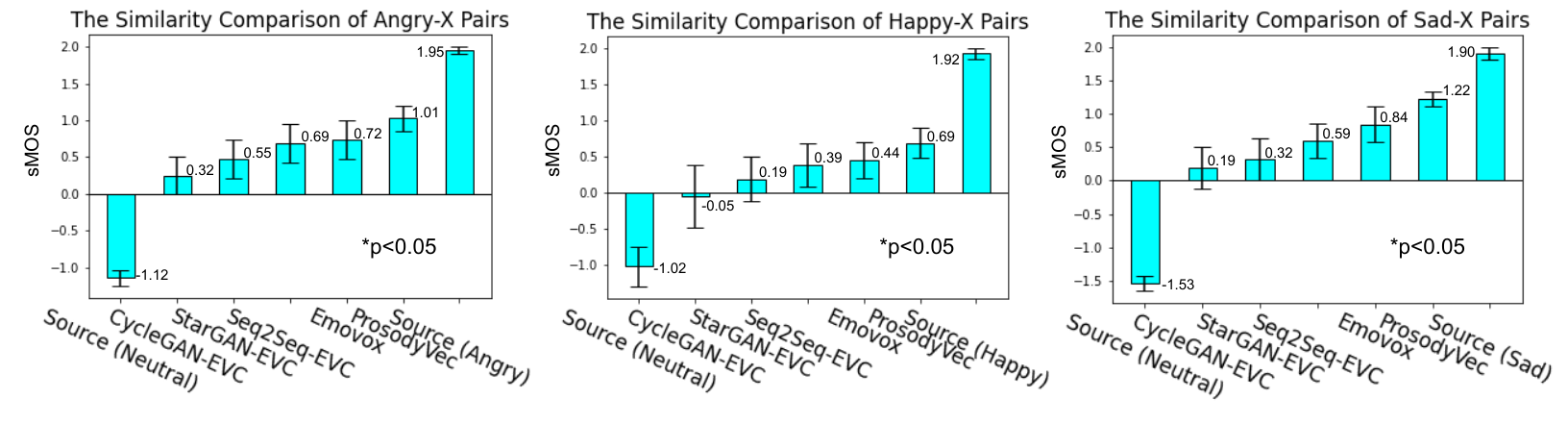}
\caption{Subjective evaluation on emotion similarity, where X is the original audio or the audio generated by the EVC models listed in the x-axis.}
\label{mos}
\vspace{-0.5cm}
\end{figure*}

\begin{figure}[h]
\setlength{\abovecaptionskip}{0cm}
\setlength{\belowcaptionskip}{-0cm}
\centering 
\small
\includegraphics[width=0.3\textwidth]{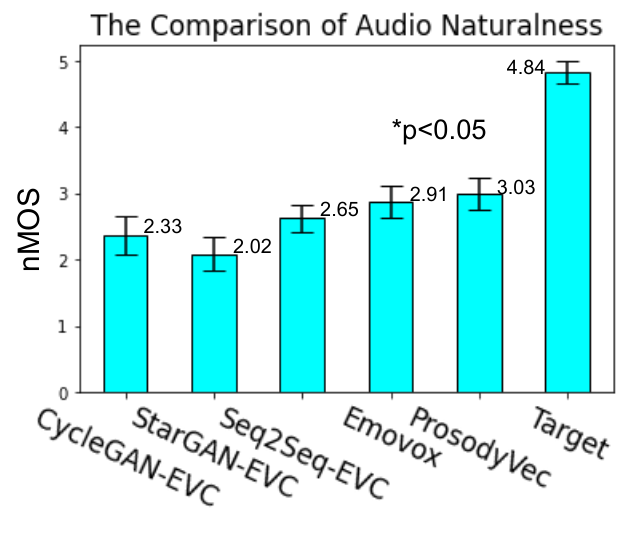}
\caption{Subjective evaluation on target and generated audio naturalness.}
\label{mos-naturaless}
\vspace{-0.3cm}
\end{figure}

We compare our model performance with supervised methods, i.e. CNN-ELM+STC attention, Auido$_{25}$~\cite{wu2021emotion}, co-attention-based fusion~\cite{zou2022speech}, IS09-classification~\cite{tarantino2019self}, TCN+self-attention w/AT~\cite{zhao2021self} and self-supervised methods, i.e. Wav2Vec~\cite{schneider2019wav2vec}, modified-CPC~\cite{riviere2020unsupervised}, DeCoAR ~\cite{ling2020decoar}, Data2Vec~\cite{baevski2022data2vec} and WavLM~\cite{chen2022wavlm}. We present the leave-one-session-out results in Table~\ref{table:leave-one-session}. Prosody2Vec achieves competitive results with some supervised models and is superior to the state-of-the-art model Wav2LM when fused with HuBERT-Large, since Prosody2Vec captures more efficient long-term variances on prosody.

For a fair comparison, we retrain SpeechFlow~\cite{qian2020unsupervised} and SpeechSplit2.0~\cite{chan2022speechsplit2} on the datasets used for Prosody2Vec pretraining. We then fine-tune the rhythm and pitch encoders for SER tasks. As shown in Table.~\ref{table:leave-one-session}, the results using the disentangled prosody representations from  SpeechFlow and SpeechSplit2.0 are not good as Prosody2Vec, since only rhythm and pitch information are decoupled. 

\begin{table}[h]
\centering
\caption{Results of SER on the IEMOCAP dataset with 5-fold cross-validation and leave-one-session-out settings.}
\resizebox{8cm}{!}{
\begin{tabular}{ccc}
\toprule
\textbf{Methods} & \textbf{WA} & \textbf{UA}\\\hline
\multicolumn{1}{l}{\textbf{Supervised Methods}}&&\\\hline
Audio$_{25}$~\cite{wu2021emotion}&60.64 & 61.32\\
IS09-classification~\cite{tarantino2019self} & 68.10 & 63.80\\
Co-attention-based fusion~\cite{zou2022speech} & 69.80 & 71.05 \\
TCN+self-attention w/AT~\cite{zhao2021self} & 65.00&66.10  \\
MTL~\cite{yue2022multi}&68.29&70.82\\
SpeechFormer++~\cite{chen2023speechformer}& 70.50 &71.50 \\\hline
\multicolumn{1}{l}{\textbf{Unsupervised/Self-supervised Methods}}&&\\\hline
Wav2Vec~\cite{schneider2019wav2vec} & 59.79&-  \\
DeCoAR 2.0~\cite{ling2020decoar} & 62.47&-  \\
Data2Vec Large~\cite{baevski2022data2vec} & 66.31&- \\
WavLM Large~\cite{chen2022wavlm} & 70.62&- \\
SpeechFlow~\cite{qian2020unsupervised} &60.43& 61.27\\
SpeechSplit2.0~\cite{chan2022speechsplit2} &62.03& 62.96\\\hline
\multicolumn{1}{l}{\textbf{Our Methods}} &     \\\hline
 Prosody2Vec & 64.10& 65.32 \\
 Prosody2Vec + HuBERT Large& \textbf{72.42} &\textbf{73.25}\\
\bottomrule
\end{tabular}}
\label{table:leave-one-session}
\vspace{-0.2cm}
\end{table}

\begin{table}[h]
\vspace{-0.2cm}
\centering
\caption{Results of SER on the IEMOCAP dataset with 10-fold cross-validation and leave-one-speaker-out settings.}
\resizebox{7cm}{!}{
\begin{tabular}{ccc}
\toprule
\textbf{Methods} & \textbf{WA} & \textbf{UA}\\\hline
\multicolumn{1}{l}{\textbf{Supervised Methods}}&&\\\hline
Attention-BLSTM-CTC~\cite{zhao2019attention}& 69.00&67.00\\
HNSD~\cite{cao2021hierarchical}& 70.50 & 72.50\\
Attention pooling~\cite{li2018attention}&71.75&68.06\\
TFCNN+DenseCap+ELM~\cite{liu2020speech}&70.34&70.78\\
CNN+GRU+SeqCap~\cite{wu2021speech}& 72.73&59.71\\
LIGHT-SERNET~\cite{aftab2022light}&70.23&70.76\\\hline
\multicolumn{1}{l}{\textbf{Self-supervised Methods}}&&\\\hline
Wav2Vec large~\cite{wang2021fine}&70.99&-\\
HuBERT large~\cite{wang2021fine}&73.01&-\\
SpeechFlow~\cite{qian2020unsupervised} &61.51& 63.25\\
SpeechSplit2.0~\cite{chan2022speechsplit2} &63.11& 63.88\\\hline
\multicolumn{1}{l}{\textbf{Our Methods}}&&\\\hline
 Prosody2Vec & 66.03& 66.57\\
 Prosody2Vec + HuBERT Large& \textbf{73.74} &\textbf{73.93}\\
\bottomrule
\end{tabular}}
\label{table:leave-one-speaker}
\vspace{-0.4cm}
\end{table}

As shown in Table~\ref{table:leave-one-speaker}, we compare our model with previous supervised and self-supervised work using leave-one-speaker-out settings. 
The self-supervised models (Wav2Vec 2.0 and HuBERT large) are first pretrained on a 60k hours speech dataset, then perform SER on IEMOCAP by partially fine-tuned.
The results of WA and UA further verify that Prosody2Vec is complementary and beneficial for the performance in widely used speech pretraining models.

\begin{figure*}
\setlength{\abovecaptionskip}{0cm}
\setlength{\belowcaptionskip}{-0cm}
\centering 
\small
\includegraphics[width=1\textwidth]{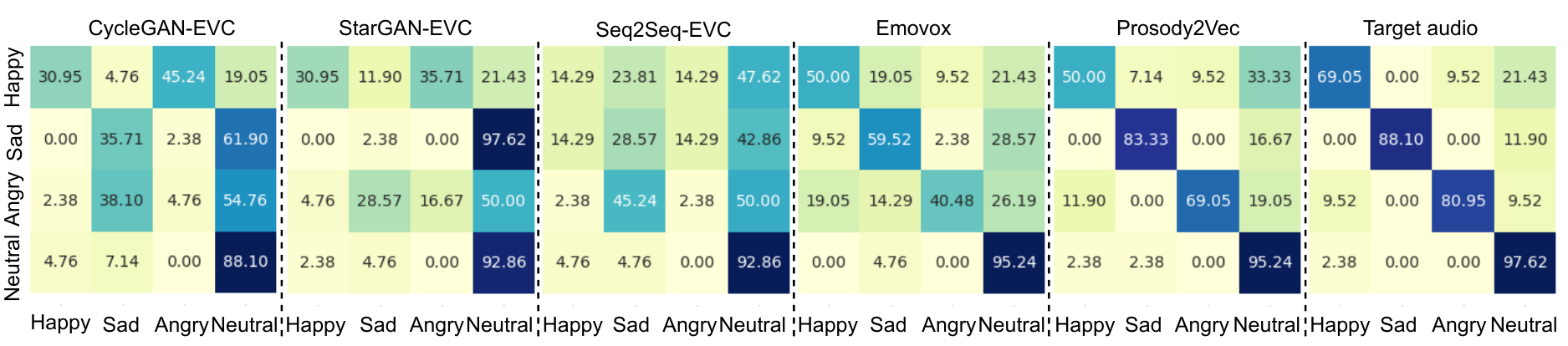}
\caption{The confusion matrices of human recognition on the converted audio by different methods. X-axis: classified labels. Y-axis: actual labels.}
\label{confusion-matrix}
\vspace{-0.3cm}
\end{figure*}

\subsection{Experiments of Emotional Voice Conversion}
\subsubsection{Experimental Setups} We follow the setups used in Emovox~\cite{zhou2022emotion} and conduct emotion conversion with the following three conditions, neutral to angry, neutral to happy, and neutral to sad. The official split of the dataset is utilized. It is worth noting that, in contrast to previous work that trains the model only on one male speaker (003), e.g. Emovox, we perform multi-speaker EVC in one model. After fine-tuning on the ESD dataset with a fixed LR of 1e-5, emotion conversion can be performed by directly replacing the input audio with expected emotions for the prosody encoder.


\subsubsection{Evaluation Metrics} The Mean Opinion Score (MOS) is utilized to subjectively evaluate the similarity between the generated and original audio. In addition, we use two objective metrics to measure the converted speech quality, i.e. Mel-cepstral distortion (MCD) and Root Mean Squared Error for F0 (F0-RMSE). 

\begin{itemize}
    \item \textbf{sMOS} is similarity MOS that is a subjective metric evaluated by the human auditory sense. For a fair comparison, the audio selection is according to the samples provided by Emovox\footnote{https://kunzhou9646.github.io/Emovox\_demo/}. The sMOS results are evaluated by 14 subjects consisting of 6 females and 8 males with ages ranging from 23 to 34 years. During testing, all 14 subjects are assigned to listen to the original audio first, followed again by the original or a generated one. Then the subjects rate the emotional similarity of the two audios with an opinion score in the range of $-2$ to $+2$ ($-2$: absolutely different, $-1$: different, $0$: cannot tell, $+1$: similar, $+2$: absolutely similar).
    \item \textbf{nMOS} is naturalness MOS which is judged on a scale of 1 (bad) to 5 (excellent).
    \item \textbf{MCD} is adopted to quantify the distortion between two mel-scale cepstral features objectively, and smaller values equal better performance. 
\begin{equation}
\setlength{\abovedisplayskip}{1pt}
\setlength{\belowdisplayskip}{1pt}
MCD=(10/\ln10)\sqrt{2\sum_{i=1}^{24}(M_i^t-M_i^c)^2}
\label{eq:mcd}
\end{equation}

\noindent where $M_i^t$ is the mel-cepstral of target emotion and $M_i^c$ is the mel-cepstral of converted audio by Prosody2Vec. 

    \item \textbf{F0-RMSE} is utilized to evaluate the distortion of frequency contour objectively.

\begin{equation}
\setlength{\abovedisplayskip}{1pt}
\setlength{\belowdisplayskip}{1pt}
F0\mbox{-}RMSE=\sqrt{\frac{1}{n}\sum_{i=1}^{N}(F_i^t-F_i^c)^2}
\label{eq:rmse}
\end{equation}

\noindent where $F_i^t$ and $F_c^t$ represent the $F0$ of target emotions and converted audio respectively. It is worth noting that we calculate the $F0$ values of the entire utterance, which includes both voiced and unvoiced regions since unvoiced segments can convey emotions as well.

\end{itemize}

\begin{table*}[!t]
\renewcommand\arraystretch{1.2}
\setlength{\abovecaptionskip}{0.cm}
\setlength{\belowcaptionskip}{-0.cm}
	\caption{Objective evaluation of emotional voice conversion with the metric of MCD and F0-RMSE}
	\label{tab:objective}
	\vspace{0.8mm}
	\centering
	\small
	\setlength\tabcolsep{8pt}
\begin{tabular}{cccccc}
\toprule
 Model & CycleGAN-EVC~\cite{zhou2020transforming} & StarGAN-EVC~\cite{rizos2020stargan} & Seq2Seq-EVC~\cite{zhou2021limited} & Emovox~\cite{zhou2022emotion} & Prosody2Vec (Ours) \\\hline
 MCD (dB) $\downarrow$ & 7.67 & 8.06 & 7.89 & \textbf{6.38} & 6.81 \\\hline
F0-RMSE (Hz) $\downarrow$ & 52.54 & 56.22 & 55.51 &  \textbf{52.23}& 53.31\\
\bottomrule
\end{tabular}
\vspace{-0.5cm}
\end{table*}

\begin{figure*}
\setlength{\abovecaptionskip}{0cm}
\setlength{\belowcaptionskip}{-0cm}
\centering 
\small
\includegraphics[width=0.85\textwidth]{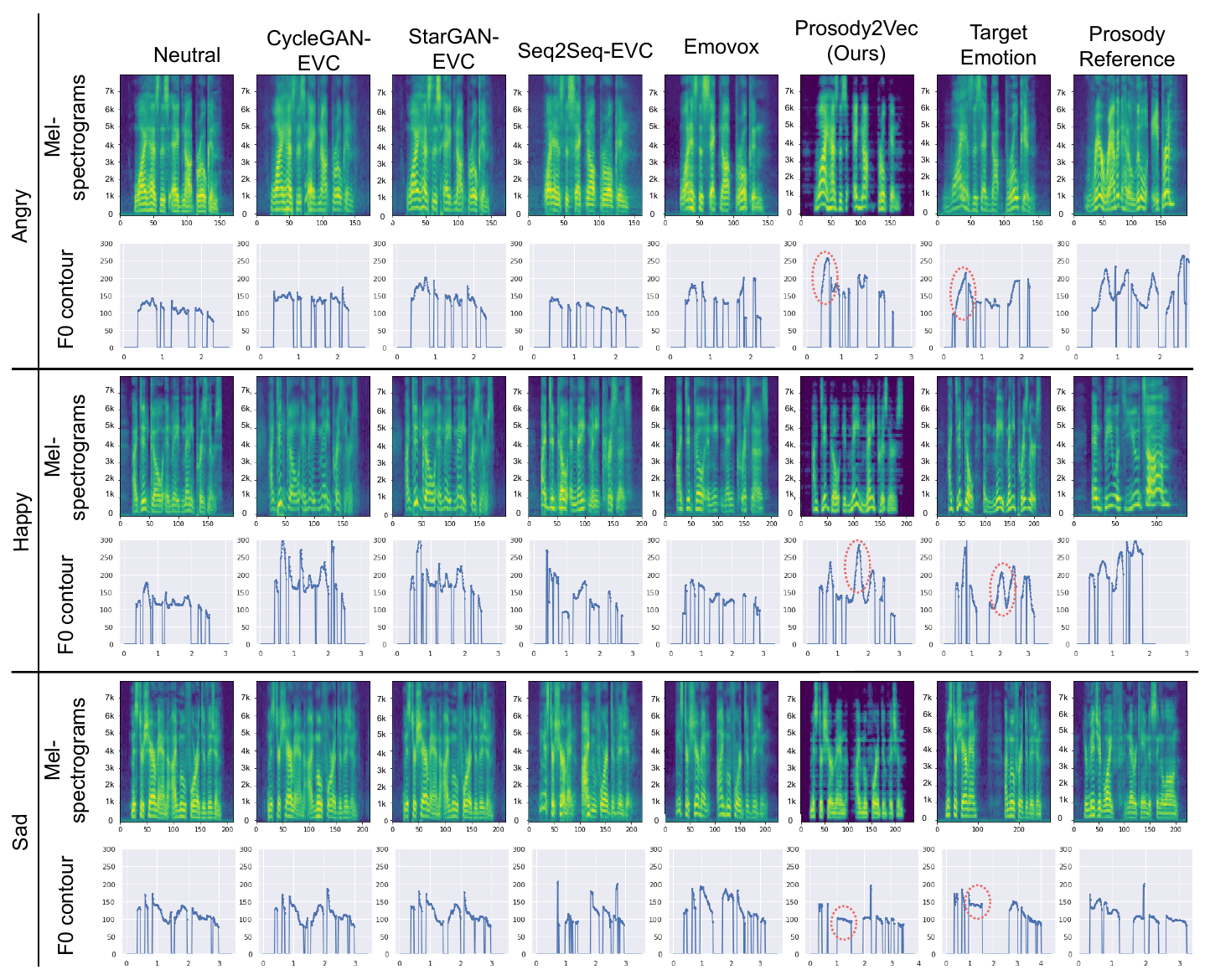}
\caption{Comparison of Mel-spectrograms and F0 contours generated by different methods for angry, happy and sad emotion conversion. Red dotted ellipses highlight F0 changes to showcase the similarity between the mel-spectrograms generated by our method and the ground truth.}
\label{mel}
\vspace{-0.5cm}
\end{figure*}

\subsubsection{Subjective Results of EVC} We compare our method with four baseline models, i.e. CycleGAN-EVC~\cite{zhou2020transforming}, StarGAN-EVC~\cite{rizos2020stargan}, Seq2Seq-EVC~\cite{zhou2021limited}, and Emovox. Fig.~\ref{mos} shows the results of sMOS regarding emotion similarity. The subjects can obviously discriminate the original emotional speech and neutral emotion with a minus score of around $-1$, as shown in the first bar in each subfigure. In addition, it is also easy to recognize the original emotional pairs with a score of around $2$, as shown in the last bars. Moreover, Prosody2Vec obtains higher scores than baselines, which reveals that our method can smoothly transfer emotional prosody into source audio.
In addition, the naturalness of generated audio by different methods is reported in Fig.~\ref{mos-naturaless}. Our method achieves competitive naturalness compared with previous work. However, it is still not so good as the target audio.

To further assess the model performance, we also ask the subjects to recognize the emotion type from a given set (neural, happy, sad, and angry) during subjective testing.  Fig.~\ref{confusion-matrix} presents the confusion matrices for each method, the darker the color, the higher the accuracy. Prosody2Vec outperforms the four baseline models with a higher accuracy in all three conversion cases.

\subsubsection{Objective Results of EVC} As shown in Table~\ref{tab:objective}, Emovox achieves the best results in both metrics. Prosody2Vec performs slightly worse than Emovox. 
 By comparing the converted audio with the target audio, we found that the audio generated by our model sounds more emotional and more expressive with different intonations or stresses. Most importantly, the rhythm and syllable duration are changed significantly. The phenomenon can be observed in the Fig. 5, where the duration of generated audio is obviously shorter than the original one.
 However, both MCD and F0-RMSE metrics are calculated frame-by-frame, the changes on duration have an important influence on the results, which leads to a slightly worse objective result by our method.

We visualize one sample for each emotion class with mel-spectrograms and F0 contours, where we transfer the expected emotional prosody from the reference prosody, as shown in Fig.~\ref{mel}. Our method generates rich variations in formants and F0 contours in comparison to the baselines.

\section{Ablation Study}
\label{sec:ablation}

In this section, we conduct a series of ablation studies to deeply understand the model architecture.

\subsection{Ablation Study on Speech Units}

We conduct ablation studies to verify the effectiveness of the deduplication process on prosody information filter with a vocabulary size of 200 units.
Specifically, we train several models based on the ECAPA-TDNN architecture but with different input features. The models are evaluated on the speech emotion recognition task.

\begin{table}[h]
\vspace{-0.2cm}
\centering
\caption{Comparison of SER results using different inputs.}
\resizebox{4.5cm}{!}{
\begin{tabular}{cc}
\toprule
\textbf{Inputs} & \textbf{WA}$\uparrow$\\\hline
Audio&55.7\\
Text&45.8\\
Duplicated Units&51.3\\
Deduplicated Units&46.2\\
\bottomrule
\end{tabular}}
\label{table:ablation-unit}
\vspace{-0.3cm}
\end{table}

As shown in Table~\ref{table:ablation-unit}, the model trained with audio inputs achieves better performance than the one trained with the text modality, since audio modality contains not only semantic content but also prosody information that is crucial for emotion recognition. 
The ``Duplicated Units" and ``Deduplicated Units" represent the unit sequence with and without repetitions respectively. The model trained with duplicated units obtains higher WA than using text. However, the deduplication units after removing repetitions achieve similar results to the one using text (45.8\% VS 46.2\%), which reveals that the deduplication process can effectively eliminate prosodic information from speech.

\subsection{Ablation Study on U2V}

We conduct ablation studies to examine the effect of BiLSTM in the U2V module.
Model performance is evaluated on the SER task by adding one additional FC layer
on top of the prosody encoder for classification. 
For a fair comparison, only the weights in the FC layer are updated while the prosody encoders are frozen.
As shown in Table~\ref{table:ablation-u2v}, the WA grows with the dimension of the BiLSTM layer.
We finally choose 256 dimensions for BiLSTM to trade off the model performance and computational costs.


\begin{table}[h]
\vspace{-0.2cm}
\centering
\caption{Ablation Study of the U2V module on SER experiments.}
\resizebox{5cm}{!}{
\begin{tabular}{ccc}
\toprule
\textbf{Module} & \textbf{Dimension} & \textbf{WA}$\uparrow$\\\hline
\multirow{3}{*}{BiLSTM}  &128&57.5\\
 &256&59.8\\
 &512&60.0\\
\bottomrule
\end{tabular}}
\label{table:ablation-u2v}
\vspace{-0.3cm}
\end{table}

\subsection{Ablation Study on Prosody Encoder}

To examine to what extent the semantics and speaker information leak from the prosody encoder, we conduct semantics and speaker probing experiments.

\subsubsection{Semantics Probing}

the semantics probing experiments are conducted on the RAVDESS~\cite{livingstone2018ryerson} dataset 
which is an emotional speech dataset recorded by 24 actors and contains 1440 utterances. 
RAVDESS consists of two kinds of semantic contents, i.e. A\textit{-``Kids are talking by the door" and }B\textit{-``Dogs are sitting by the door"}.

We first generate speech samples by controlling the inputs of the unit and prosody encoders with different combinations, for example, AB means feeding utterances with semantics A into the unit encoder of a pretrained Prosody2Vec model while feeding inputs B into the prosody encoder. We then use the Whisper~\cite{radford2022robust} ASR system to transcribe the generated speech signals into text transcriptions.
Finally, The sentence-level accuracy of being recognized as A, B, or X is calculated. X is neither A nor B, which is caused by word errors in the results of the Whisper system.

As shown in Table~\ref{table:ablation-semantics}, the transcribed texts are consistent with the prosody encoder input with high accuracy (98.8\% and 99.2\%), 
which suggests that the semantics of the generated speech is controlled by the unit encoder and no linguistic content is leaked from the prosody encoder.

\begin{table}[h]
\vspace{-0.2cm}
\centering
\caption{Semantics probing on the unit and prosody encoder inputs.}
\resizebox{6cm}{!}{
\begin{tabular}{cc|ccc}
\toprule
\multicolumn{2}{c|}{Input pair} & \multicolumn{3}{c}{Acc.$\uparrow$} \\\hline
Unit     & Prosody              & A         & B         & X       \\\hline
A              & B              & 98.8      & 0.0       & 1.2     \\
B              & A              & 0.0       & 99.2      & 0.8    \\
\bottomrule

\end{tabular}}
\label{table:ablation-semantics}
\end{table}

\subsubsection{Speaker Probing}

we found that the decoder may learn speaker information from the prosody encoder even if speaker embeddings are provided.
Therefore, we force the prosody encoder to learn only prosody-related information by constraining the dimension of prosody representations.
We train Prosody2Vec with different dimensions of the prosody and the unit encoder, as shown in Table~\ref{table:ablation-prosody-dim}.
We conduct speaker verification and SER experiments
to examine the residual speaker information in prosody and unit representations.

\begin{itemize}
    \item \textbf{Speaker Verification}. 
Speaker verification is conducted on the LibriSpeech subset \texttt{train-clean-100} ~\cite{panayotov2015librispeech} which is randomly split into training and testing sets with the ratio of $9:1$ from 251 speakers. The prosody encoder is frozen and one FC layer is added on top of it to perform classification and maintain the learned prosodic knowledge.
As shown in Table~\ref{table:ablation-prosody-dim}, when the dimension of prosody representation is set to 64, we obtain the lowest speaker verification accuracy (28.0\%). However, the accuracy increases to 66.7\% when the dimension grows to 320, which reveals that constraining the dimension of prosody representations can effectively mitigate speaker information leak from the prosody encoder. In comparison, the vector dimension has a minor impact on the unit encoder.

\begin{table}[h]
\vspace{-0.2cm}
\centering
\caption{Ablation study on the dimension of prosody and unit representations.}
\resizebox{7cm}{!}{
\begin{tabular}{c|cc|cc}
\toprule
          & \multicolumn{2}{c}{SV (Acc.)} & \multicolumn{2}{c}{SER (WA)} \\\hline
Dimension & prosody      & unit      & prosody     & unit     \\\hline
64        & 28.0         & 18.3      & 63.2        & 57.3     \\
128       & 30.1         & 18.1      & 63.6        & 57.5     \\
192       & 34.5         & 18.7      & 64.1        & 59.3     \\
256       & 56.1         & 18.7      & 63.4        & 59.8     \\
320       & 66.7         & 19.2      & 61.1        & 59.4   \\
\bottomrule
\end{tabular}}
\label{table:ablation-prosody-dim}
\vspace{-0.3cm}
\end{table}

\item \textbf{Speech Emotion Recognition}. 
In addition, we also examine the effect of unit and prosody dimensions on emotion recognition. 
The SER experiments are conducted with the leave-one-session settings on the IEMOCAP dataset. 
Similar to the speaker verification experiments, one additional FC layer is added on top of the frozen prosody encoder.
As shown in Table~\ref{table:ablation-prosody-dim}, the WA goes up and then down as the dimension of prosody representations increases.
The speaker verification experiments reveal that high dimensions cause speaker information leaks, 
which leads to the poor generalization of prosody representations and degrades the SER performance. As the unit dimension increases, the SER accuracy also experiences a slight rise. 
We finally report the SER and EVC results with 192-dimensional prosody and 256-dimensional unit representations respectively in Section IV. EXPERIMENT to trade-off the performance of SER and speaker verification. 

\end{itemize}

\begin{figure*}
\setlength{\abovecaptionskip}{0cm}
\setlength{\belowcaptionskip}{-0cm}
\centering 
\small
\includegraphics[width=0.7\textwidth]{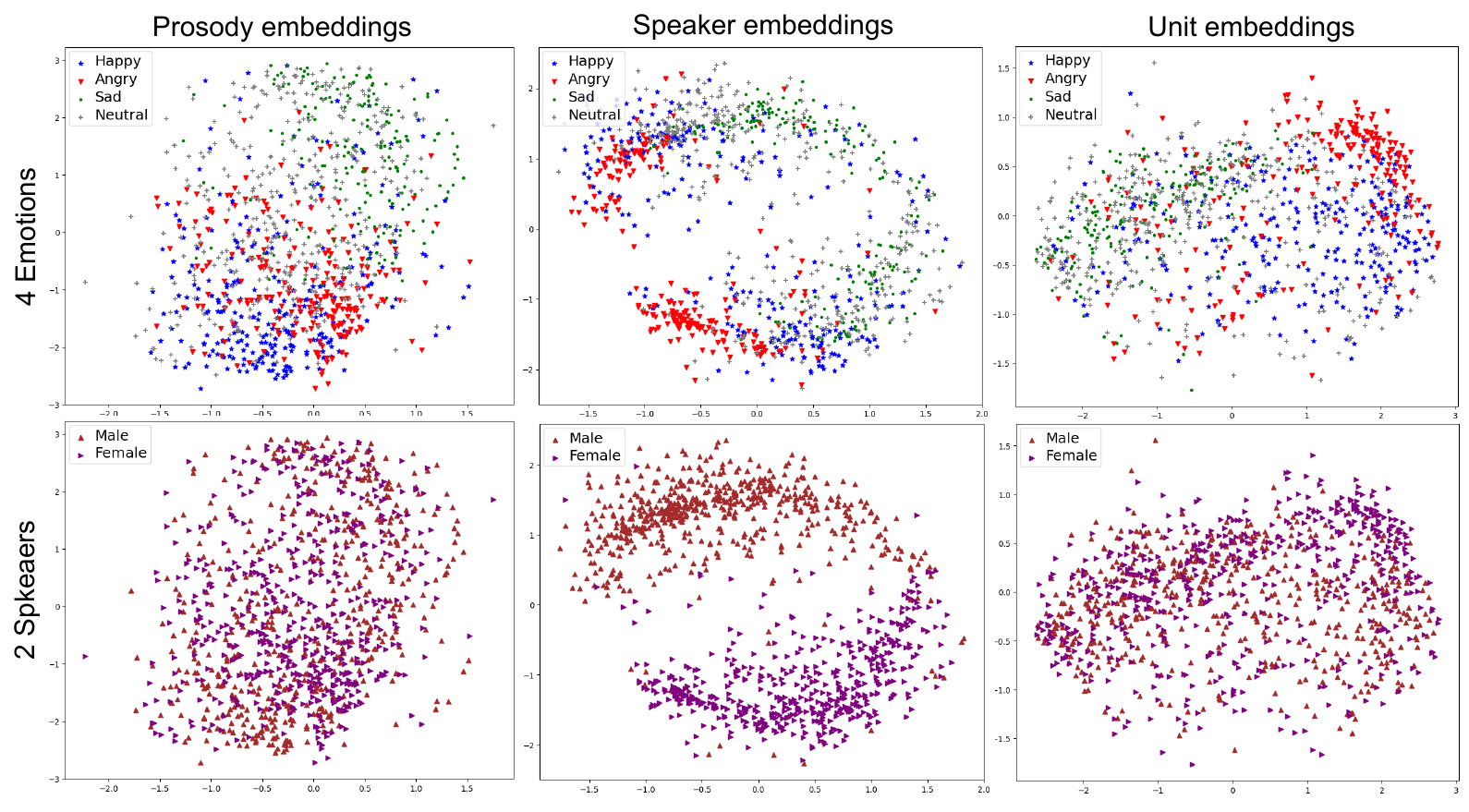}
\caption{Visualization of prosody, speaker, and unit embeddings from pretrained models colored with 4 emotion labels and 2 speaker labels on the IEMOCAP dataset.}
\label{tnse}
\vspace{-0.5cm}
\end{figure*}

\subsection{Embedding visualization}
To further straightforwardly understand Prosody2Vec, we visualize the prosody, speaker, and unit embeddings learned by the three encoders with t-SNE~\cite{van2008visualizing}.
We choose the audio samples in the first session of IEMOCAP uttered by two speakers with 4 emotions, i.e. angry, happy, sad, and neural.
The sentence-level unit embeddings are obtained by averaging on the time domain.
It is noteworthy that all embeddings are extracted with the pretrained Prosody2Vec without fine-tuning on the IEMOCAP dataset.
As we can observe in Fig.~\ref{tnse}, we color the embeddings in the emotion and speaker dimensions to explore their representation ability on emotion and speaker classifications.

\begin{figure*}
\setlength{\abovecaptionskip}{0cm}
\setlength{\belowcaptionskip}{-0cm}
\centering 
\small
\includegraphics[width=0.7\textwidth]{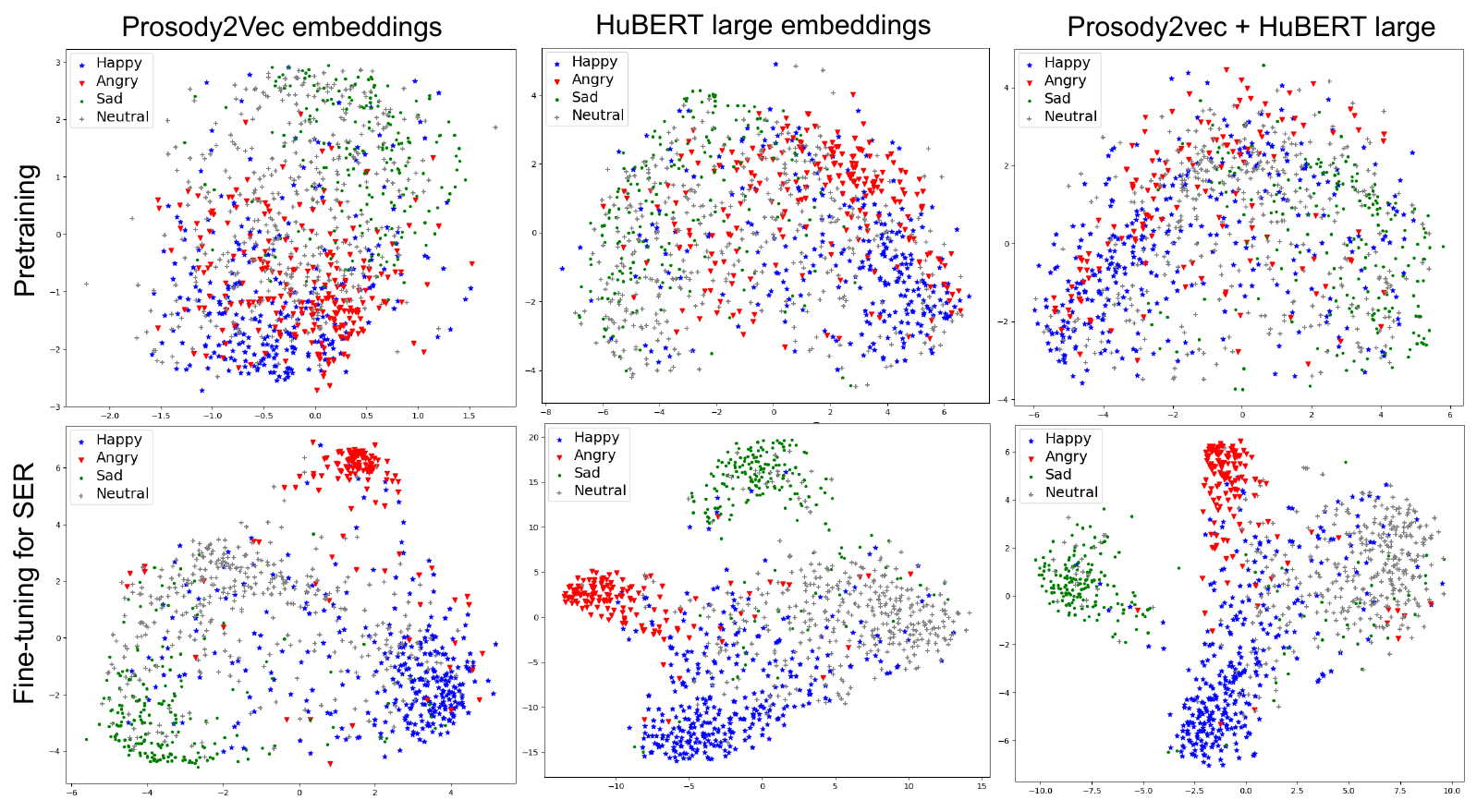}
\caption{Comparison of pretrained and fine-tuned embeddings colored with 4 emotions on the IEMOCAP dataset.}
\label{tnse2}
\vspace{-0.5cm}
\end{figure*}

For a fair comparison, the representations presented in Fig.~\ref{tnse} come from the pretrained Prosody2Vec which is not fine-tuned on emotional datasets, since the unit and speaker encoders will not be fine-tuned on downstream tasks. This is the reason why Fig.~\ref{tnse} does not show separated clusters on the emotion domains. We found the same phenomenon in the HuBERT model. As shown in Fig.~\ref{tnse2}, the first row is the visualization of the representations from the pretrained models, and the second row shows the model outputs after fine-tuning on emotion classification tasks. 
We can conclude that although the representations extracted from the pretrained models cannot distinguish emotions, they demonstrate great potential when fine-tuning on domain-specific datasets. Moreover, the visualizations also reveal that Prosody2Vec synergistically integrates with the semantic representation model HuBERT. This harmonious integration results in a noticeably enhanced performance. 

\begin{figure}[h]
\vspace{-0.2cm}
\setlength{\abovecaptionskip}{0cm}
\setlength{\belowcaptionskip}{-0cm}
\centering 
\small
\includegraphics[width=0.5\textwidth]{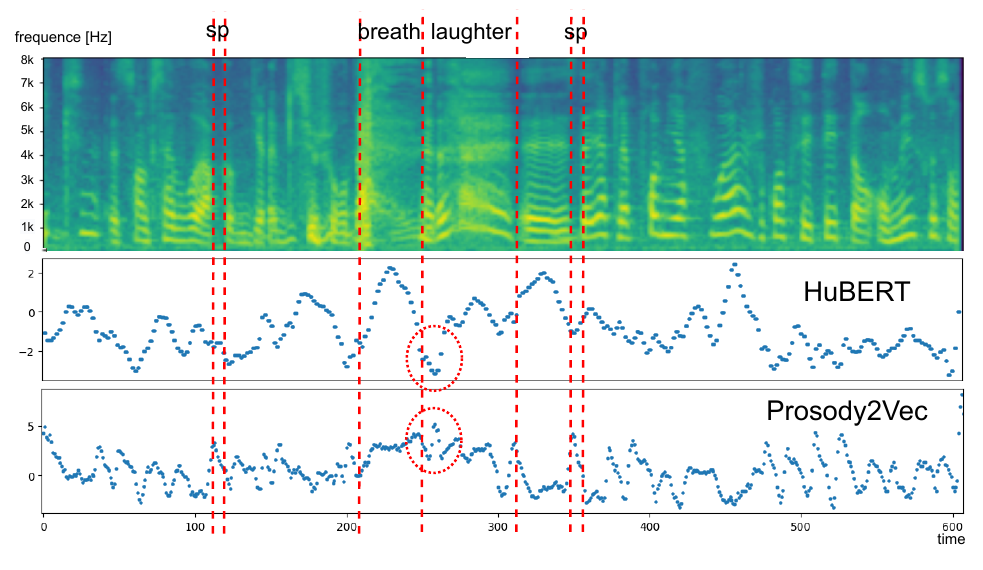}
\caption{Visualization of HuBERT and Prosody2Vec representations after dimension reduction with PCA in the time domain, where sp is short for short pause. The red dotted ellipses highlight different activation.}
\label{pca}
\vspace{-0.3cm}
\end{figure}

Furthermore, to facilitate an intuitive comparison, we employ Principal Component Analysis (PCA) to reduce the frame-level HuBERT representations (1024 dimensions) and the outputs from the attentive pooling layer in Prosody2Vec (3072 dimensions) into a single dimension, as shown in Fig.~\ref{pca}. The audio sample is spoken with breath and laughter, conveying a sense of happy emotion. Compared with HuBERT, Prosody2Vec can better represent the timing information, such as short pauses between words and the durations of segments. In addition, Prosody2Vec has a different activation on non-verbal areas, for instance breath and laughter (highlighted with red circles).

\section{Potential Applications and Discussion}
\label{discussion}

\begin{figure*}[!th]
\setlength{\abovecaptionskip}{0cm}
\setlength{\belowcaptionskip}{-0cm}
\centering 
\small
\includegraphics[width=0.9\textwidth]{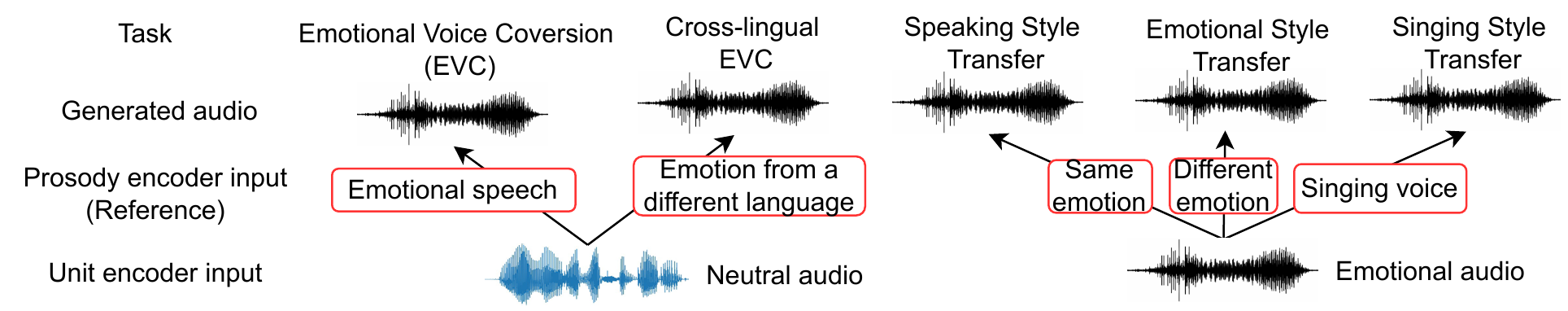}
\caption{Overview of five generation tasks presented in this paper when changing the inputs of the prosody encoder and the unit encoder.}
\label{task}
\vspace{-0.5cm}
\end{figure*}

We have shown that our proposed Prosody2Vec can capture utterance-level prosody information, which significantly boosts the performance of SER and EVC tasks. As shown in Fig.~\ref{task}, we discuss some potential applications of our model on cross-lingual EVC and speaking, emotional, and singing style transfer. Cross-lingual EVC transfers an emotional style from a different language to the source language.
Singing style transfer refers to transforming speaking prosody into a given melody.
Speaking style transfer intends to change prosodic attributes, for instance, stress position and intensity level in the generated audio, while keeping the emotion type unchanged. 
Emotional style transfer aims to convert one emotion to a different one, for example, angry to happy.  
We only present cross-lingual EVC and singing style transfer in this section. Speaking and emotional style transfer are discussed in Appendix A and B respectively.
It is worth noting that all potential applications are conducted without any fine-tuning with task-related datasets. 
Lastly, we conclude by discussing the benefits and limitations of Prosody2Vec.

\subsection{Cross-lingual Emotional Voice Conversion}
\begin{figure}[h]
\setlength{\abovecaptionskip}{0cm}
\setlength{\belowcaptionskip}{-0cm}
\centering 
\small
\includegraphics[width=0.35\textwidth]{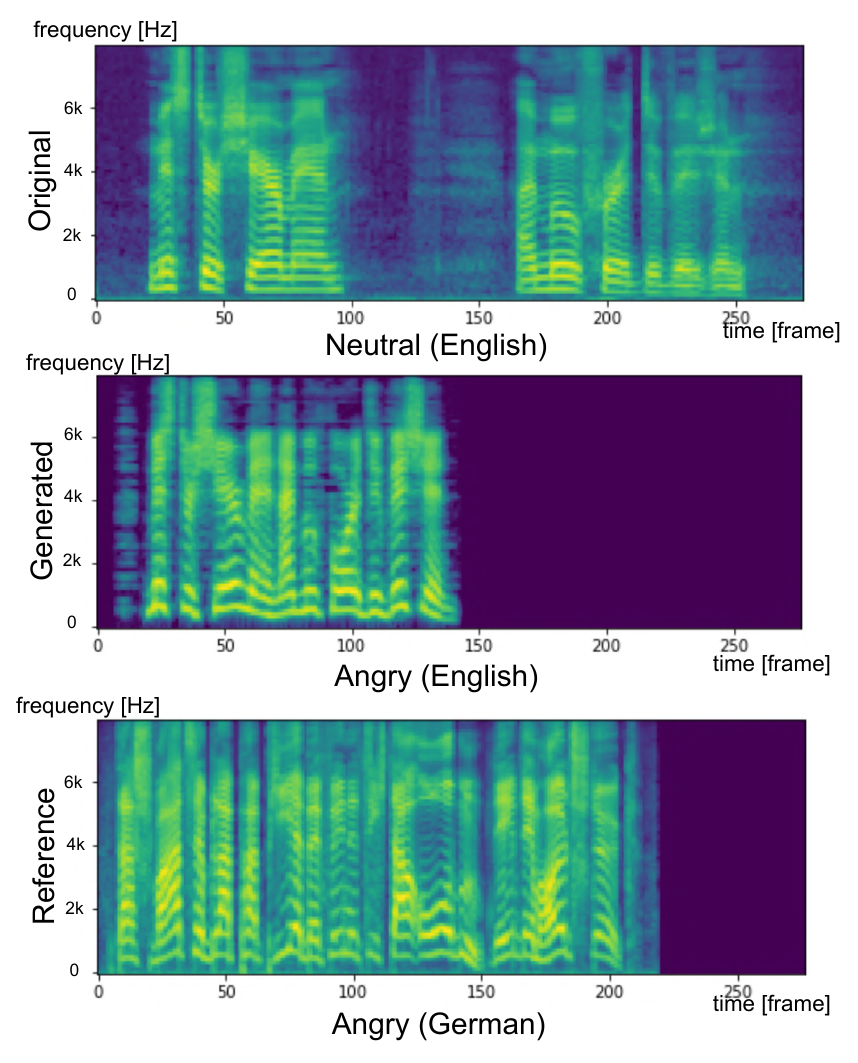}
\caption{Cross-lingual emotional voice conversion (German to English). For a convenient comparison on rhythm and tempo, we pad short sequences to the length of the original audio.}
\label{cross-language}
\vspace{-0.2cm}
\end{figure}

We found that Prosody2Vec can perform zero-shot cross-lingual emotional style transfer. As shown in Fig.~\ref{cross-language}, we convert an English neutral utterance into another emotion (angry) by transferring the prosodic information from a German reference. We only use English data for pretraining and the model never sees any German speech.

Compared to the original English neutral audio, the given German reference is uttered with a relatively fast tempo. As we can see in the second picture of Fig.~\ref{cross-language}, Prosody2Vec successfully transfers the tight rhythm in an unseen German reference into the English utterance but keeps semantic content invariant. The middle short pause in the original audio is even removed to perform a rapid tempo.

\subsection{Singing Style Transfer}

\begin{figure}[h]
\setlength{\abovecaptionskip}{0cm}
\setlength{\belowcaptionskip}{-0cm}
\centering 
\small
\includegraphics[width=0.35\textwidth]{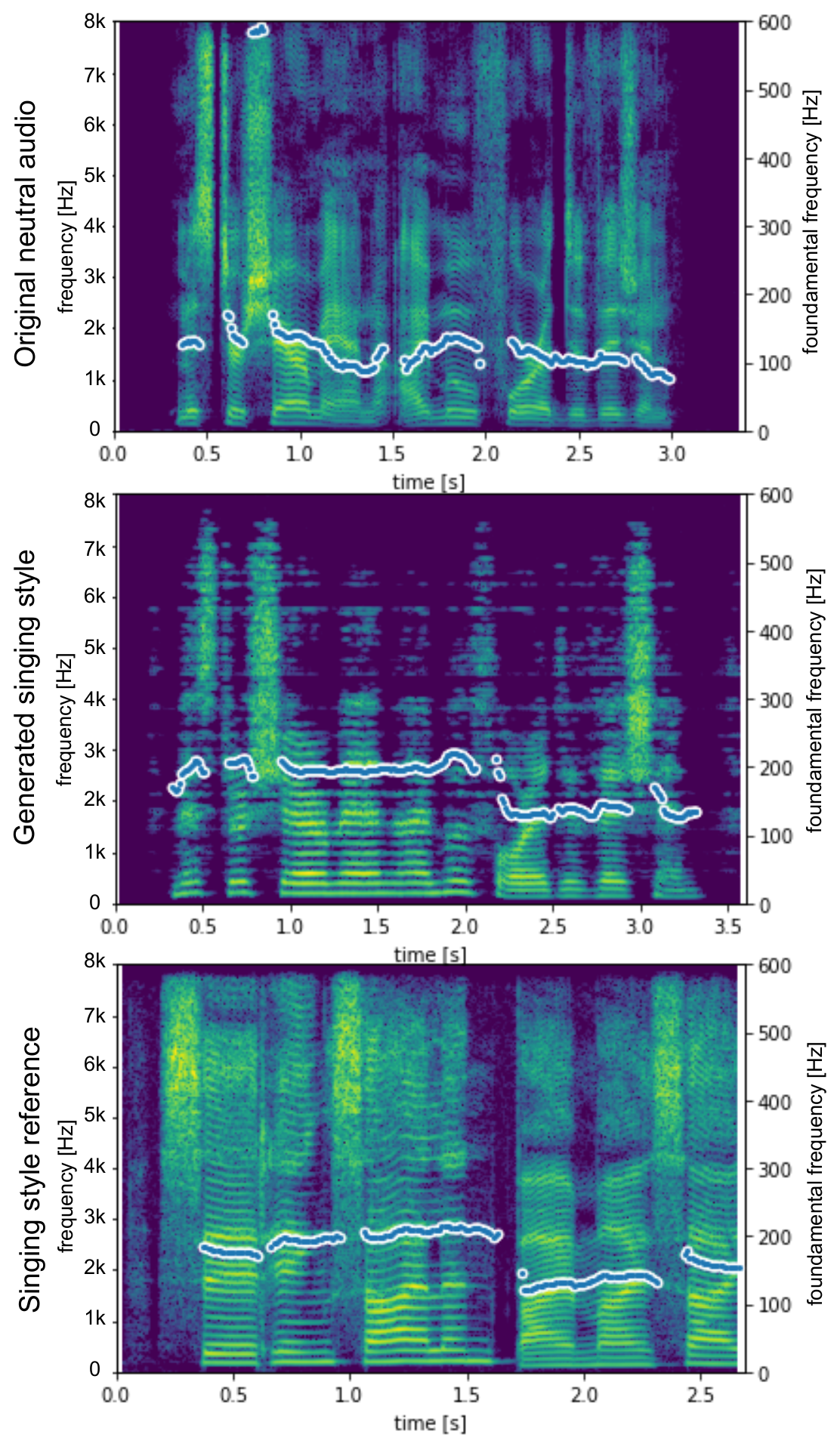}
\caption{Transferring the melody from a singing voice to a spoken utterance, where pitch is marked in white color.}
\label{sing-transfer}
\vspace{-0.3cm}
\end{figure}

We visualize the pitch with Parselmouth\footnote{https://parselmouth.readthedocs.io/en/stable/} in each mel-spectrogram since the spoken intonation and the musical melody are highly related to pitch variance. 
From Fig.~\ref{sing-transfer} we can see when feeding a singing voice to the prosody encoder, Prosody2Vec can successfully transfer the melody in the given reference into the source utterance, which suggests that Prosody2Vec can be used for music synthesis or style conversion.

\subsection{Discussion}

The style transfer tasks shown above further reveal that our proposed model successfully disentangles prosodic information which is independent of semantic content and robust to unseen styles and languages. We highly recommend listening to the audio samples on our demo website\footnote{https://leyuanqu.github.io/Prosody2Vec}.

However, we found that the speech quality for emotion conversion is damaged sometimes. For example, the distortions around 2000Hz in the second picture of Fig.~\ref{sing-transfer}. The generated distortions will have a noticeable effect during listening. This is mainly because, during training, the model always receives inputs belonging to the same speaker. It is difficult for the model to only focus on prosodic information when directly replacing the prosody encoder input with a different speaker since the model has never seen such combinations during training. 

Moreover, a surprising finding is that when we randomly replace a few unit values with random numbers or remove a few k-means clustering units in the input sequences, the quality or semantics of the generated audio are only slightly influenced, which reveals that the discrete units are very robust compared to the text sequences transcribed by ASR systems. Hence, it is worth further exploring our system under more challenging conditions, such as speech with noise or reverberation.

\section{Conclusion and Future Work}
\label{conclusion}
In this paper, we propose Prosody2Vec to learn emotional prosody representations from speech, which consists of three encoders: a unit encoder to transform speech signals into discrete units, a speaker encoder to provide speaker identity information, a prosody encoder to extract utterance-level representations, and a TTS-based decoder to reconstruct mel-spectrograms by relying on the aforementioned three information flows. Only the weights of the prosody encoder and the decoder are trainable in order to force the prosody encoder to capture prosodic changes when minimizing the distance between generated and original speech signals. Prosody2Vec relies neither on paired audio nor on any emotion or prosody labels. The experimental results on SER and EVC reveal that the Prosody2Vec structure learns efficient prosodic features which achieve considerable improvements compared to the state-of-the-art models for emotion classification and emotion transfer.



The current model is trained only for English which is a non-tonal language. It is worth verifying our methods on some tonal languages, e.g. Mandarin and Thai. Furthermore, since emotional expressions are highly influenced by languages and cultures, it would be interesting to investigate the prosodic patterns and mechanisms across languages. One major reason limiting the performance of modern SER systems is the lack of large-scale and high-quality emotional corpora. Augmenting emotional speech data using Prosody2Vec with EVC would be a promising approach.

\small

\bibliographystyle{IEEEbib}
\bibliography{main}


\appendix

 In addition to cross-lingual EVC and sing style transfer, 
 we show more applications of Prosody2Vec in the Appendix.

\subsection{Speaking Style Transfer}

\begin{figure}[h]
\setlength{\abovecaptionskip}{0cm}
\setlength{\belowcaptionskip}{-0cm}
\centering 
\small
\includegraphics[width=0.35\textwidth]{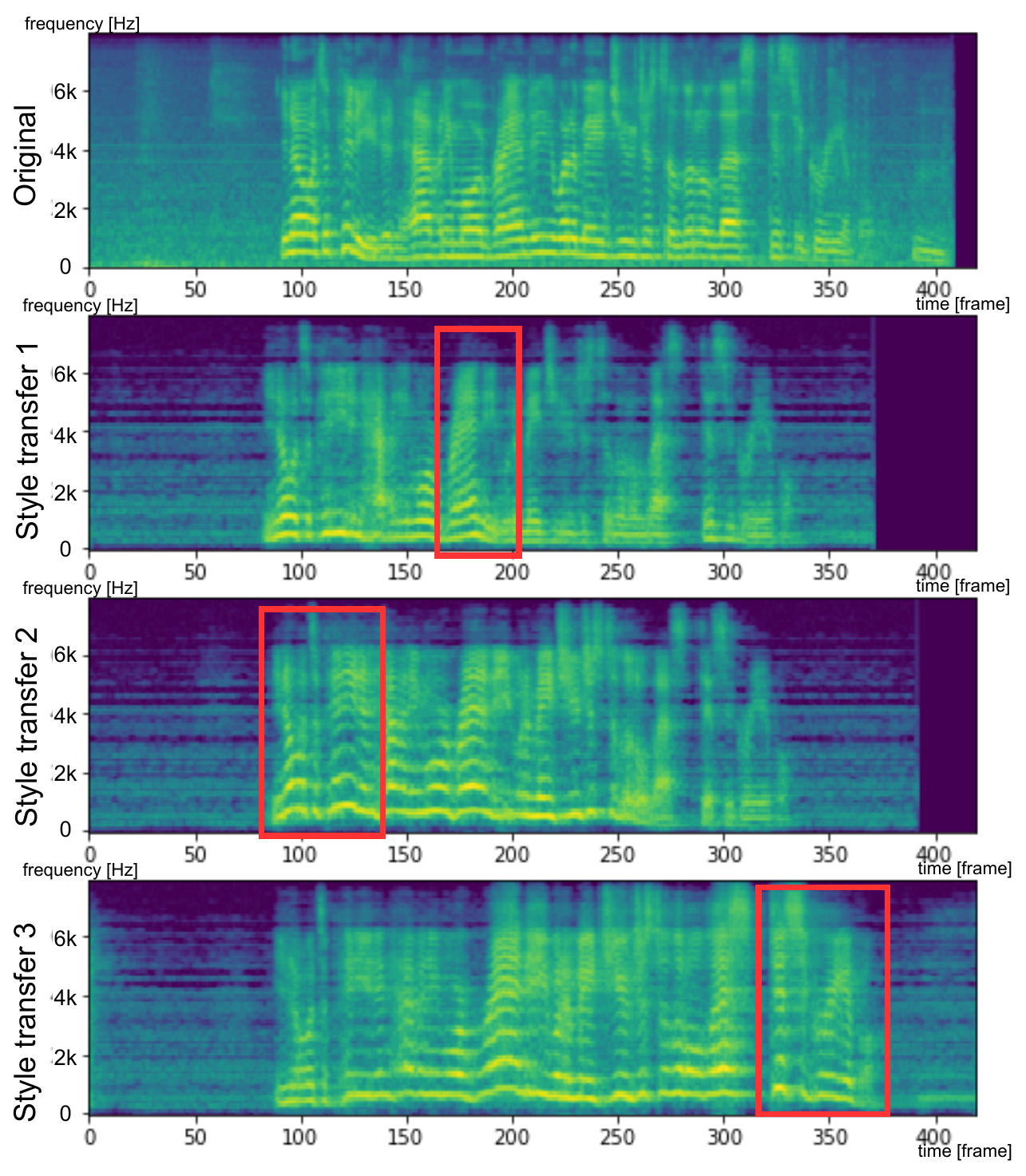}
\caption{Speaking style transfer for the utterance of ``You know, it's a pity you didn't have any more brandy. It would have made you just a little less disagreeable".}
\label{data-augmentation}
\vspace{-0.5cm}
\end{figure}

When feeding a reference with a given emotion type into the prosody encoder, we found that the model generates emotional audio with different stress positions or intonations. In addition, a different emotional intensity can arise through conversion. We compare the original and generated mel-spectrograms in Fig.~\ref{data-augmentation}, in which some differences on stress to demonstrate the transferred styles are highlighted with red boxes. This application can be used to augment emotional speech to mitigate class imbalance and data scarcity problems in the SER task.

\subsection{Emotional Style Transfer}

\begin{figure}[h]
\setlength{\abovecaptionskip}{0cm}
\setlength{\belowcaptionskip}{-0cm}
\centering 
\small
\includegraphics[width=0.35\textwidth]{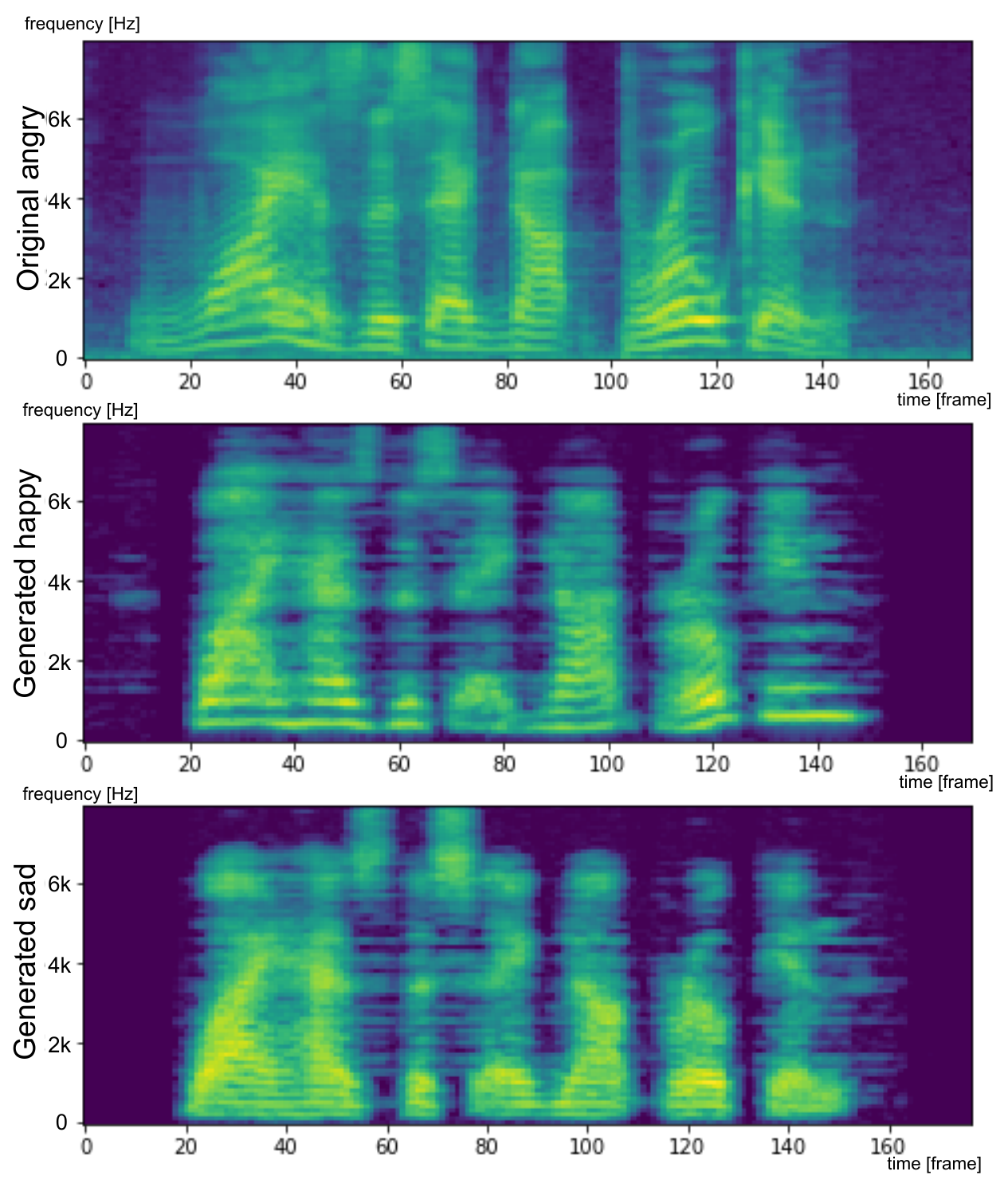}
\caption{Samples of converting angry emotion to happy and sad ones.}
\label{emotion-transfer}
\vspace{-0.5cm}
\end{figure}

Inspired by the fact that humans can easily manipulate emotion expressions while not altering the semantic content~\cite{oster2010using}, here we show that emotion expressions are independent of semantics from a signal processing perspective. As shown in Fig.~\ref{emotion-transfer}, the original utterance ``Why is this egg not broken?" is uttered with angry emotion. We can smoothly convert the source audio to happy or sad emotions while retaining semantic information.

\begin{IEEEbiography}[{\includegraphics[height=1.25in,clip,keepaspectratio]{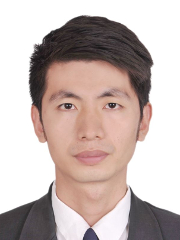}}]{Leyuan Qu} received the M.Sc degree in computer science from Beijing Language and Culture University, Beijing, China, in 2017. He received the Ph.D. degree from the Department of Informatics, University of Hamburg, Hamburg, Germany, in 2021. He is currently a Postdoctoral Fellow with Zhejiang Lab, Hangzhou, China. His main research interests include speech representation learning, multi-modal learning, affective computing and self-supervised learning.
\end{IEEEbiography}

\vspace{-30 pt}

\begin{IEEEbiography}[{\includegraphics[height=1.25in,clip,keepaspectratio]{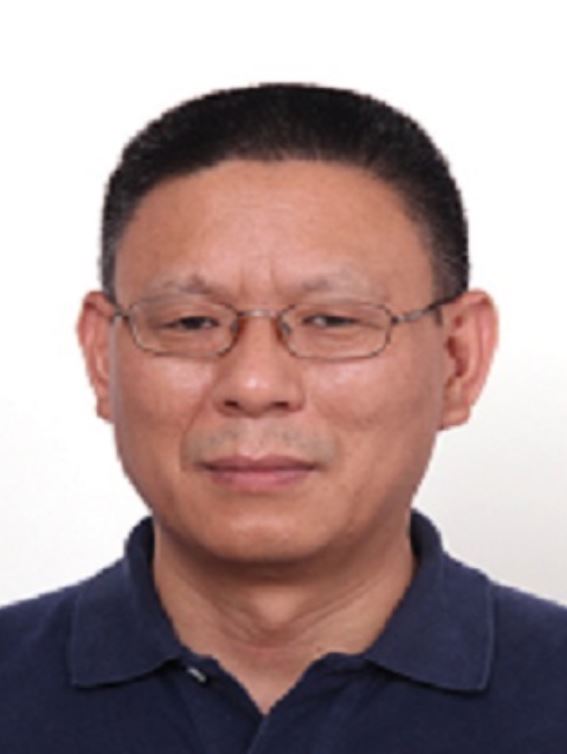}}]{Taihao Li} received his Ph.D degree in Information Science and Systems Engineering from National University of Tokushima in 2006. He was a researcher in Harvard University from 2006 till 2011. He was a principle scientist in Flatley Discovery Lab from 2011 till 2019. He is now a Senior Research Expert and Deputy Director of Cross-Media Intelligence Research Center in Zhejiang Lab, Hangzhou, China. He has published more than 30 related papers in well-known journals and conferences around topics like affective computing, image processing and multi-modal information fusion, etc. He also hosted or participated in 18 projects in the United States, Japan and China and has applied more than 30 patents for multi-modal emotion recognition.
\end{IEEEbiography}

\vspace{-30 pt}

\begin{IEEEbiography}[{\includegraphics[height=1.25in,clip,keepaspectratio]{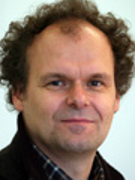}}]{Cornelius Weber} received the Diploma degree in physics from the University of Bielefeld, Bielefeld, Germany, and the Ph.D. degree in computer science from the Technische Universität Berlin, Berlin, Germany, in 2000. He was a Post-Doctoral Fellow of Brain and Cognitive Sciences with the University of Rochester, Rochester, NY, USA. From 2002 to 2005, he was a Research Scientist of Hybrid Intelligent Systems with the University of Sunderland, Sunderland, U.K. He was a Junior Fellow with the Frankfurt Institute for Advanced Studies, Frankfurt am Main, Germany, until 2010. He is now a Laboratory Manager with the Knowledge Technology group, Universität Hamburg, Hamburg, Germany. His current research interests include computational neuroscience with a focus on vision, unsupervised learning, and reinforcement learning.
\end{IEEEbiography}

\vspace{-30 pt}

\begin{IEEEbiography}[{\includegraphics[height=1.25in,clip,keepaspectratio]{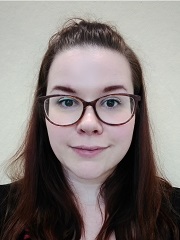}}]{Theresa Pekarek-Rosin} received her M.Sc. degree in computer science from the University of Hamburg in 2021.
She is currently a Ph.D. student and Research Associate with the Knowledge Technology Group, University of Hamburg, Germany.
Her main research interests include speech recognition of non-standard speech, representation learning, and continual learning of speech characteristics.
\end{IEEEbiography}

\vspace{-30 pt}

\begin{IEEEbiography}[{\includegraphics[height=1.25in,clip,keepaspectratio]{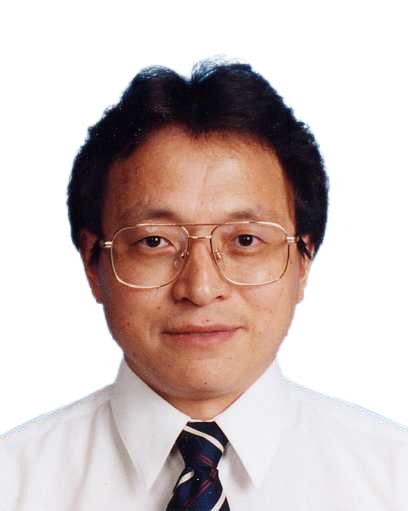}}]{Fuji Ren} received his Ph. D. degree in 1991 from the Faculty of Engineering, Hokkaido University, Japan. From 1991 to 1994, he worked at CSK as a chief researcher. In 1994, he joined the Faculty of Information Sciences, Hiroshima City University, as an Associate Professor. Since 2001, he has been a Professor of the Faculty of Engineering, Tokushima University. He is a Chair Professor of University of Electronic Science and Technology of China from 2022. His current research interests include Natural Language Processing, Artificial Intelligence, Affective Computing, Emotional Robot. He is the Academician of The Engineering Academy of Japan and EU Academy of Sciences. He is a senior member of IEEE, Editor-in-Chief of International Journal of Advanced Intelligence, a vice president of CAAI, and a Fellow of The Japan Federation of Engineering Societies, a Fellow of IEICE, a Fellow of CAAI. He is the President of International Advanced Information Institute, Japan.
\end{IEEEbiography}

\vspace{-30 pt}

\begin{IEEEbiography}[{\includegraphics[height=1.25in,clip,keepaspectratio]{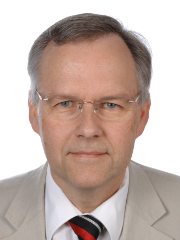}}]{Stefan Wermter} is Full Professor at the University of Hamburg, Germany, and Director of the Knowledge Technology Institute in the Dept. of Informatics. He has previously held positions at the University of Dortmund,  University of Massachusetts, the International Computer Science Institute in Berkeley and the University of Sunderland. His main research interests are in neural networks, hybrid knowledge technology, neuroscience-inspired computing, cognitive robotics, natural language processing and human-robot interaction. He has been associate editor of the journal `IEEE Transactions on Neural Networks and Learning Systems' and he is on the advisory board of `Connection Science' and `International Journal for Hybrid Intelligent Systems' and he is on the editorial board of the journals `Cognitive Computation', `Neurosymbolic Artificial Intelligence' and `Journal of Computational Intelligence'. He is coordinator of the international doctoral training network TRAIL, co-coordinator of the international collaborative research centre on Crossmodal Learning (TRR-169) and he currently serves as the President of the European Neural Network Society.
\end{IEEEbiography}

\vfill

\end{document}